%% file: p.tex
\documentclass[sigconf,screen,nonacm]{sty/acmart}

\AtBeginDocument{%
  }

\include{pkgs}

\newcommand{\sys}{\mbox{\textsc{AFuzz}}\xspace}

\input{cmds}

\begin{document}

\input{hdr}
\date{}
\input{abstract}
\maketitle

\sloppy

\input{intro}
\input{motivation}

\input{design}
\input{impl}
\input{eval}
\input{problem}
\input{discussion}
\input{relwk}
\input{conclusion}

\footnotesize
\setlength{\bibsep}{3pt}
\bibliographystyle{sty/ACM-Reference-Format}
\bibliography{p,sslab,conf}

\normalsize

\input{appendix}

\end{document}

%% file: pkgs.tex
\usepackage{subcaption}
\usepackage{xspace,fancyvrb,multirow}
\usepackage{array,underscore,relsize}
\usepackage{enumitem}

\usepackage{fp}
\usepackage{siunitx}

\usepackage{minted}

\usepackage{algorithm}
\usepackage{algpseudocode}

\usepackage{pifont}
\usepackage{colortbl}

%% file: cmds.tex
\newcommand{\cc}[1]{\mbox{\smaller[0.5]\texttt{#1}}}

\fvset{fontsize=\footnotesize,xleftmargin=8pt,numbers=left,numbersep=5pt}

\input{code/fmt}

\def\Snospace~{\S{}}

\newif\ifdraft\drafttrue
\newif\ifnotes\notestrue
\ifdraft\else\notesfalse\fi

\input{glyphtounicode}
\newcolumntype{R}[1]{>{\raggedleft\let\newline\\\arraybackslash\hspace{0pt}}p{#1}}

\newcommand{\squishlist}{
\begin{itemize}[noitemsep,nolistsep,leftmargin=*]
  \setlength{\itemsep}{-0pt}
}
\newcommand{\squishend}{
  \end{itemize}
}

\usepackage{tikz}
\usetikzlibrary{tikzmark,arrows.meta}
\newcommand*\WC[1]{%
\begin{tikzpicture}[baseline=(C.base)]
\node[draw,circle,inner sep=0.2pt](C) {#1};
\end{tikzpicture}}

\newcommand{\PP}[1]{
\noindent{\bf \IfEndWith{#1}{.}{#1}{#1.}}
}

\newcommand{\boxbeg}{
\vspace{2px}
\noindent\begin{tabular}{|l|}\hline
\begin{minipage}{3.2in}
\vspace{2px}
\noindent
}

\newcommand{\boxend}{
\vspace{2px}
\end{minipage}\\ \hline
\end{tabular}
\vspace{-10pt}
}

\definecolor{redcellcolor}{RGB}{244,199,195}
\definecolor{greencellcolor}{RGB}{183,225,205}
\definecolor{yellowcellcolor}{RGB}{252,232,178}

\newcommand{\etal}{et al.}

\newcommand{\appendixref}{\hyperref[s:appendix]{Appendix}\xspace}

\newcommand{\Analyzer}{\textsc{Analyzer}\xspace}
\newcommand{\Investigator}{\textsc{Investigator}\xspace}
\newcommand{\ScenarioAnalyzer}{\textsc{Scenario Analyzer}\xspace}
\newcommand{\Validator}{\textsc{Validator}\xspace}
\newcommand{\BugFinder}{\textsc{Bug Finder}\xspace}

\newcommand{\seedtotal}{3,146\xspace}
\newcommand{\seedeval}{669\xspace}

\newcommand{\bugfixed}{28\xspace}
\newcommand{\bugconfirmed}{two\xspace}
\newcommand{\bugduplicate}{three\xspace}
\newcommand{\bugreview}{seven\xspace}

\newcommand{\bugtotal}{40\xspace}

\newcommand{\bugexperimental}{six\xspace}

\newcommand{\bidI}{480788626}
\newcommand{\bidII}{481074858}
\newcommand{\bidIII}{484300742}
\newcommand{\bidIV}{484789568}
\newcommand{\bidV}{485281841}
\newcommand{\bidVI}{496807872}
\newcommand{\bidVII}{499659070}
\newcommand{\bidVIII}{502030575}
\newcommand{\bidIX}{482083327}
\newcommand{\bidX}{482199449}
\newcommand{\bidXI}{489494031}
\newcommand{\bidXII}{491881374}
\newcommand{\bidXIII}{498010834}
\newcommand{\bidXIV}{498904291}
\newcommand{\bidXV}{498904295}
\newcommand{\bidXVI}{498904299}
\newcommand{\bidXVII}{499018901}
\newcommand{\bidXVIII}{499150024}
\newcommand{\bidXIX}{499155349}
\newcommand{\bidXX}{499188872}
\newcommand{\bidXXI}{499206651}
\newcommand{\bidXXII}{499254994}
\newcommand{\bidXXIII}{499254996}
\newcommand{\bidXXIV}{499323105}
\newcommand{\bidXXV}{499520013}
\newcommand{\bidXXVI}{499659062}
\newcommand{\bidXXVII}{499672315}
\newcommand{\bidXXVIII}{499834467}
\newcommand{\bidXXIX}{500224598}
\newcommand{\bidXXX}{500507435}
\newcommand{\bidXXXI}{500507436}
\newcommand{\bidXXXII}{500536164}
\newcommand{\bidXXXIII}{502035871}
\newcommand{\bidXXXIV}{502083475}
\newcommand{\bidXXXV}{482083331}
\newcommand{\bidXXXVI}{482088900}
\newcommand{\bidXXXVII}{499027764}
\newcommand{\bidXXXVIII}{499103615}
\newcommand{\bidXXXIX}{499150035}
\newcommand{\bidXL}{499155348}

\newcommand{\smbidI}{2032677}
\newcommand{\smbidII}{2032761}
\newcommand{\smbidIII}{2032821}
\newcommand{\smbidIV}{2032819}
\newcommand{\smbidV}{2032943}

\newcommand{\jscbidI}{312632}
\newcommand{\jscbidII}{312664}
\newcommand{\jscbidIII}{312667}
\newcommand{\jscbidIV}{312669}
\newcommand{\jscbidV}{312672}
\newcommand{\jscbidVI}{312675}
\newcommand{\jscbidVII}{312681}
\newcommand{\jscbidVIII}{312682}
\newcommand{\jscbidIX}{312684}
\newcommand{\jscbidX}{312685}
\newcommand{\jscbidXI}{312687}
\newcommand{\jscbidXII}{312688}
\newcommand{\jscbidXIII}{312689}
\newcommand{\jscbidXIV}{312690}

\newcommand{\cveII}{CVE-2026-2649\xspace}
\newcommand{\cveVIII}{CVE-2026-7902\xspace}

%% file: hdr.tex
\title{Agentic Fuzzing: Opportunities and Challenges}

\ifdefined\DRAFT
 \pagestyle{fancyplain}
 \lhead{Rev.~\therev}
 \rhead{\thedate}
 \cfoot{\thepage\ of \pageref{LastPage}}
\fi

\author{Junyoung Park}
\email{parkjuny@kaist.ac.kr}
\affiliation{
    \institution{KAIST}
    \city{Daejeon}
    \country{Republic of Korea}
}

\author{Insu Yun}
\email{insuyun@kaist.ac.kr}
\affiliation{
    \institution{KAIST}
    \city{Daejeon}
    \country{Republic of Korea}
}

%% file: abstract.tex
\begin{abstract}
    Fuzzers and static analyzers find many bugs but struggle with logic bugs in mature codebases. Triggering such a bug often requires multi-step reasoning that produces no distinctive execution feedback, and variants can appear across implementations too different for a single pattern to match. Recent LLM-assisted approaches help, but they use LLMs as auxiliaries rather than as the reasoning engine.
    
    We propose \emph{agentic fuzzing}, a bug-finding approach seeded by historical bugs in which deep agents perform the reasoning directly. Given a reference bug, the agent analyzes its root cause, hypothesizes new scenarios elsewhere in the codebase that may share that cause, and verifies each hypothesis by generating and running proof-of-concept code. This lets the agent find variants that differ completely in trigger path or code structure from the reference.

    We identify three practical challenges in implementing agentic fuzzing: harness engineering, redundant investigations across seeds with similar root causes, and scheduling seeds in a large corpus. We address these in \sys through a four-stage agent pipeline, \emph{scenario coverage} that deduplicates previously explored scenarios, and a DPP-MAP scheduler that orders seeds by diversity. We ran \sys on the V8 JavaScript engine for about one month, finding \bugtotal bugs (including \bugduplicate duplicates), receiving a total \$35{,}000 bounty, and being assigned two CVEs. \sys also found 19 bugs (including one duplicate) in SpiderMonkey and JavaScriptCore using the seeds from V8. However, agentic fuzzing is in its early stages with several remaining open problems we discuss in the paper. Still, we think it points to a promising direction for finding logic bugs.
\end{abstract}

%% file: intro.tex
\section{Introduction}
\label{s:intro}

As software systems grow in size and complexity, automated bug detection techniques are commonly used to maintain software quality and security. Among them, fuzzing~\cite{libfuzzer,clusterfuzz,gross2023fuzzilli,wachter2025dumpling} and static analysis~\cite{codeql,kim2017vuddy,woo2022movery,xiao2020mvp,kang2022tracer,feng2024fire,jang2012redebug,xu2025enhancing} have been widely adopted for their ability to discover bugs at scale. Fuzzers discover bugs by generating a large number of diverse inputs and observing execution feedback such as code coverage or crashes. Static analysis matches predefined patterns or traces data flows across the program without executing it. These techniques have discovered tens of thousands of bugs in projects like Chromium~\cite{clusterfuzz} and are widely deployed across open-source projects~\cite{codeql}.

However, these techniques struggle to find \emph{logic bugs} in mature codebases. A blog post by the V8 team~\cite{v8sandbox} revealed that V8 JavaScript engine vulnerabilities are rarely memory corruption bugs, but instead logic bugs that have the potential to turn into ones when exploited. Because logic bugs do not necessarily exhibit observable signals such as crashes, fuzzers are likely to miss them since they rely on such signals to guide their search. Static analyzers also cannot reliably find them because finding such bugs usually requires reasoning about complex conditions that may not be easily captured by general patterns. For example, detecting an integer overflow bug may require static analyzers to understand how to achieve sufficiently large integer values through multiple stages of execution, which is difficult to express as a simple general pattern.

Historical bugs can provide useful guidance for finding logic bugs, as they demonstrate root causes and trigger conditions that can be used to hypothesize new vulnerability scenarios. However, existing tools lack the reasoning capability to use this guidance effectively. Fuzzers can use Proof-of-Concept (PoC) code as a seed to generate or mutate inputs, exploring around the original trigger path. However, because they do not understand the meaning of the seed, they can only explore around the original trigger path, missing similar bugs with different trigger paths (e.g., JavaScript vs. WebAssembly). Static analyzers can match bug patterns derived from historical bugs. However, a pattern broad enough to cover different code structures (e.g., different components) loses the precision needed to avoid false positives, while a more precise one does not carry over to another implementation of the same logical flaw.

Recent works use LLMs to assist fuzzing~\cite{zhang2025g2fuzz,xia2024fuzz4all,meng2024chatafl,yang2025kernelgpt,deng2023titanfuzz,deng2024fuzzgpt,yang2024whitefox,eom2024covrl,lyu2024promptfuzz,chen2025elfuzz} and static analysis~\cite{li2025iris,lekssays2025llmxcpg}. They use LLMs to generate test inputs, infer protocol grammars, or label code property graph nodes. However, in each case, LLMs are used as auxiliaries, while coverage feedback or pattern matching still steers the search. Because the deep reasoning LLMs are capable of is mostly not used to drive the search, the fundamental limitations of existing approaches persist. Logic bugs that require reasoning about complex conditions across multiple stages of execution can avoid detection, and similar bugs that differ in trigger paths or code structures are unmatched.

We propose \emph{agentic fuzzing}, a new approach that uses deep agents instead as the primary reasoning engine for finding bugs, seeded by historical bugs. Recent \emph{deep agents}~\cite{deep-agents,claudecode,codex,gemini-cli} are capable of complex, multi-step reasoning and code manipulation, matching what a human auditor is capable of. Therefore, an agentic fuzzer mimics how a human auditor finds bugs: it takes a reference bug as a seed, analyzes its root cause, hypothesizes new vulnerability scenarios that share the same root cause, and verifies each hypothesis through code reasoning and PoC generation. This lets agentic fuzzers find similar bugs even when they completely differ in trigger paths and code structures, because the agent can understand the root cause at a high level and then reason about how it may occur across the codebase.

In practice, implementing an agentic fuzzer raises three problems. First, there is no obvious way to define the agent's tasks---a problem known as \emph{harness engineering}~\cite{harness-engineering}. It requires domain expertise and iterative experimentation to find the right design. Second, when an agentic fuzzer processes multiple seeds, it often performs redundant investigations because each seed is analyzed independently and semantically similar seeds lead to overlapping hypotheses. Third, given a large seed corpus, an agentic fuzzer needs to decide which seeds to process and in what order to maximize bug discovery. Ideally, an agentic fuzzer would prioritize the most promising seeds, but estimating each seed's potential is difficult.

We implemented a prototype \sys, which addresses each of these problems as follows. First, \sys decomposes the task into a four-stage pipeline of specialized agents (\Analyzer, \Investigator, \ScenarioAnalyzer, and \Validator), each focused on a single objective. This pipeline enables deep analysis of each seed and yields the highest per-seed bug discovery rate in our evaluation. Second, \sys tracks \emph{scenario coverage}, a record of all previously explored location-hypothesis pairs, to skip redundant investigations. Third, \sys uses a strategy to disperse seed selection across diverse root causes, using a DPP-MAP~\cite{macchi1975coincidence,chen2018fast,han2020map} algorithm. This strategy allows \sys to cover a broad range of root causes early on.

We evaluated \sys on the V8 JavaScript engine~\cite{veight} for about one month, processing about 750 of the \seedtotal seeds collected from the Chromium issue tracker~\cite{chromiumissuetracker}. As a result, \sys discovered \bugtotal bugs (including \bugduplicate duplicates), of which four vulnerabilities received a total bounty of \$35{,}000 and were assigned two CVEs (\cveII and \cveVIII). The bugs are not limited to a specific component but span the parser, interpreter, and multiple compilers. As we detail in \autoref{ss:eval-newbugs}, several of these bugs are complex logic bugs that are hard to find with conventional fuzzing or static analysis.

Agentic fuzzing is still in its early stages, with several open problems. These include the high operational cost (we evaluated only about 23.8\% of our seed corpus within our budget), the assumption that reference bugs are available (which does not hold for closed-source software), and the fragility of design choices across model generations. We discuss these in \autoref{s:problem}.

Our contributions are as follows.
\begin{itemize}[leftmargin=*]
\item We propose \emph{agentic fuzzing}, a new bug-finding approach that uses deep agents as the primary reasoning engine and takes reference bugs as seeds.
\item We identify three practical challenges in agentic fuzzing: harness engineering, redundant investigations across seeds, and seed scheduling over a large seed corpus.
\item We design and implement a prototype \sys with a four-stage agent pipeline, scenario coverage, and DPP-MAP seed scheduling.
\item We evaluated \sys primarily on V8, discovering \bugtotal bugs (including \bugduplicate duplicates), receiving a \$35{,}000 bounty and two CVEs. Our case studies show that \sys can find bugs that conventional fuzzing and static analysis tools rarely find.
\item We discuss open problems for agentic fuzzing, including cost effectiveness, the need for reference bugs, and design uncertainty.
\end{itemize}

%% file: motivation.tex
\section{Motivation}
\label{s:motivation}

Using a motivating example, we demonstrate the challenges of finding logic bugs and how existing automated approaches struggle to find them.
Then, we show how \emph{agentic fuzzing} addresses these challenges using deep LLM agents and reference bugs.
Finally, we discuss the challenges in scaling this approach to large codebases with many reference bugs, which motivates our design of \sys.

\subsection{Motivating Example}
\label{ss:motiv-variant}

\autoref{fig:motiv-vuln} shows a code snippet of an integer truncation vulnerability in V8's Turboshaft compiler that \sys discovered during evaluation. The \cc{Operation} base class stores the operand count of each IR node in a \cc{uint16\_t} field \cc{input\_count} (line~\ref{line:input-count}). However, the constructor accepts \cc{input\_count} as a \cc{size\_t} (line~\ref{line:ctor-narrow}), implicitly truncating it to 16 bits during initialization. 
A \cc{DCHECK} assertion guards this truncation (line~\ref{line:dcheck}), but it is stripped in release builds, leaving the truncation unchecked. An attacker can exploit this truncation to cause out-of-bounds access when subsequent compiler phases use the truncated \cc{input\_count}. This leads to memory corruption and potential arbitrary code execution.

To trigger the vulnerability, we use WebAssembly's \cc{br\_table} instruction (i.e., switch-case), which can carry up to 65,520 entries (\cc{kV8MaxWasmFunctionBrTableSize}). When Turboshaft compiles a \cc{br\_table}, each entry adds one predecessor to the destination block. When multiple branches merge, Turboshaft creates a \cc{PhiOp} whose operand count equals the total number of predecessors. We place two \cc{br\_table} instructions in opposite arms of an if-else structure as shown in \autoref{fig:motiv-trigger}, each with 33,000 entries targeting the same merge point. This creates a \cc{PhiOp} with 66,000 predecessors, exceeding the \cc{uint16\_t} limit and triggering the truncation.

\begin{figure}[t]
    \begin{subfigure}[t]{\columnwidth}
        \input{code/motiv-vuln.cpp}
        \caption{Vulnerable code in V8's Turboshaft IR.}
        \label{fig:motiv-vuln}
        \begin{tikzpicture}[remember picture, overlay]
          \draw[-{Stealth[length=6pt,width=5pt]}, thick, red!60!black]
            ([xshift=2pt, yshift=2pt]pic cs:vinit)
            .. controls ++(14pt, 0) and ++(14pt, 0) ..
            node[pos=0.5, right=2pt, yshift=3pt, font=\sffamily\scriptsize\bfseries, text=red!60!black] {truncated}
            ([xshift=2pt, yshift=2pt]pic cs:vfld);
        \end{tikzpicture}
    \end{subfigure}
    \begin{subfigure}[t]{\columnwidth}
        \input{code/motiv-trigger.wat}
        \caption{Simplified WebAssembly structure that triggers the vulnerability.}
        \label{fig:motiv-trigger}
    \end{subfigure}
    \caption{Motivating example: an integer truncation vulnerability (\cveII) in V8's Turboshaft compiler discovered by \sys. The red arrow in (a) shows \cc{size\_t}$\to$\cc{uint16\_t} truncation.}
    \label{fig:motiv-example}
\end{figure}

\subsection{Limitations \& Key Insights}
In this section, we discuss the problem of finding logic bugs and the insights that motivate our design of \sys.

\subsubsection{Large and complex codebases}
\label{ss:large-and-complex-codebases}
It is challenging to identify logic bugs in large and complex codebases. For example, V8~\cite{veight}, the JavaScript engine used in Google Chrome, adopts a multi-phase execution pipeline consisting of parsing, an interpreter (Ignition), a baseline compiler (Sparkplug), and an optimizing compiler (TurboFan). V8 later added a mid-tier compiler, Maglev, in 2023~\cite{maglevblog}, further increasing its complexity. Each component is designed with distinct performance goals, making JavaScript code transition dynamically across multiple execution tiers at runtime. This complexity makes it difficult to reason about program behavior and detect subtle logic bugs such as the one in our motivating example.

\begin{figure}[t]
    \begin{subfigure}[t]{\columnwidth}
        \input{code/motiv-ref.cpp}
        \caption{Vulnerable code in V8's Maglev compiler.}
        \label{fig:motiv-ref}
        \begin{tikzpicture}[remember picture, overlay]
          \draw[-{Stealth[length=5pt,width=4.5pt]}, semithick, densely dashed, blue!50!black]
            ([xshift=2pt, yshift=2pt]pic cs:msrc)
            .. controls ++(24pt, 0) and ++(24pt, 0) ..
            node[pos=0.5, right=2pt, yshift=-4pt, font=\sffamily\scriptsize\bfseries, text=blue!50!black] {data flow}
            ([xshift=2pt, yshift=2pt]pic cs:mic);
          \draw[-{Stealth[length=5pt,width=4.5pt]}, semithick, densely dashed, blue!50!black]
            ([xshift=2pt, yshift=2pt]pic cs:mic)
            .. controls ++(8pt, 0) and ++(8pt, 0) ..
            ([xshift=2pt, yshift=2pt]pic cs:mparam);
          \draw[-{Stealth[length=5pt,width=4.5pt]}, semithick, densely dashed, blue!50!black]
            ([xshift=2pt, yshift=2pt]pic cs:mparam)
            .. controls ++(8pt, 0) and ++(8pt, 0) ..
            ([xshift=2pt, yshift=2pt]pic cs:mcast);
          \draw[-{Stealth[length=6pt,width=5pt]}, thick, red!60!black]
            ([xshift=2pt, yshift=2pt]pic cs:mcast)
            .. controls ++(20pt, 0) and ++(20pt, 0) ..
            node[pos=0.4, left=2pt, font=\sffamily\scriptsize\bfseries, text=red!60!black] {truncated}
            ([xshift=2pt, yshift=2pt]pic cs:mfld);
        \end{tikzpicture}
    \end{subfigure}
    \begin{subfigure}[t]{\columnwidth}
        \input{code/motiv-ref-trigger.js}
        \caption{JavaScript code that triggers the vulnerability~\cite{referencebug}.}
        \label{fig:motiv-ref-trigger}
    \end{subfigure}
    \caption{Reference vulnerability (CVE-2025-10892) in V8's Maglev compiler~\cite{referencebug}. Blue dashed arrows and the red arrow in (a) show the data flow and integer truncation, respectively.}
    \label{fig:motiv-ref-example}
\end{figure}

\PP{Insight 1: Historical bugs as hints}
Although it is difficult to discover the motivating example from scratch, historical bugs in V8 can provide useful guidance. \autoref{fig:motiv-ref} shows CVE-2025-10892 in V8's Maglev compiler~\cite{referencebug}. Notably, this bug shares a common pattern with the motivating example: a 16-bit representation of input counts, truncation from wider integer types, and language features that push the count beyond its limit, despite differences in backend and input path.

We now describe this bug in detail.
In Maglev, the \cc{NodeBase} base class packs each IR node's metadata into a 64-bit bitfield, storing the input count in a 16-bit field \cc{InputCountField} (line~\ref{line:maglev-bitfield}). When the \cc{Allocate} function creates a node, it encodes the input count into this bitfield (line~\ref{line:maglev-encode}). However, \cc{BitField::encode} accepts the input count as a \cc{size\_t} and silently truncates it to 16 bits (line~\ref{line:maglev-truncate}). 
This vulnerability is triggered through the \cc{GeneratorStore} operation, which saves all local variables when suspending a generator function. The IR graph builder computes the input count by adding the parameter count, the bytecode register count, and a fixed number of \cc{kFixedInputCount=2} (line~\ref{line:maglev-compute}). Then, the IR graph builder passes the input count to node creation as an \cc{int} (line~\ref{line:maglev-create}). When we create a generator with a sufficiently large number of local variables (e.g., 65,534), the local variables become bytecode registers, and the input count exceeds the 16-bit limit.

\subsubsection{Lack of reasoning}
\label{ss:lack-of-reasoning}
Existing approaches struggle to find bugs even with the help of historical bugs. There are two main approaches that use historical bugs: coverage-guided fuzzing and variant analysis. First, coverage-guided fuzzing uses the Proof-of-Concept (PoC) code as a seed and mutates it to find variants without understanding its meaning. Unfortunately, while the reference bug and the motivating example share a similar root cause (i.e., integer truncation in compiler IR), their PoC codes are completely different. In particular, the former uses JavaScript for creating a generator with a large number of local variables (\autoref{fig:motiv-ref-trigger}), while the latter uses WebAssembly to do so (\autoref{fig:motiv-trigger}). Therefore, coverage-guided fuzzers are unable to find such variants.

Second, variant analysis tools, such as CodeQL~\cite{codeql} or Joern~\cite{yamaguchi2014modeling}, use static analysis to find variants. These tools detect variants by matching predefined code patterns or signatures~\cite{codeql,yamaguchi2014modeling,kim2017vuddy,woo2022movery}, measuring syntactic or structural similarity~\cite{jang2012redebug,xu2025enhancing}, or analyzing program slices or traces~\cite{xiao2020mvp,kang2022tracer,feng2024fire}. However, the reference bug and the motivating example occur in completely different implementations: one uses Maglev and the other uses Turboshaft. As a result, existing variant analysis tools also fail to find such variants.

\PP{Insight 2: LLM for reasoning}
To address these challenges, we use LLMs to reason about the code. LLMs can understand code semantics and perform complex reasoning that is challenging for existing tools. In the motivating example, LLMs can abstract away low-level implementation details such as Maglev's bitfield encoding and Turboshaft's constructor, and focus on high-level semantics: they both store wide input counts in a 16-bit field. Then, LLMs can reason about how to make the input count exceed the 16-bit limit, for example by using large \cc{br\_table} instructions to create a \cc{PhiOp} with too many predecessors. As a result, LLMs can match the vulnerability pattern across Maglev and Turboshaft and figure out how to trigger the vulnerability, discovering the motivating example from the reference bug.

\subsubsection{Limited LLM integration}
Instead of using LLMs as the primary reasoning engine, existing LLM-based approaches use LLMs as auxiliaries to assist fuzzing or static analysis. This limited integration prevents them from fully leveraging LLMs' reasoning capabilities to find logic bugs like our motivating example. For example, LLM-assisted fuzzers~\cite{zhang2025g2fuzz,xia2024fuzz4all} use LLMs to generate inputs that can maximize code coverage, rather than to reason about why and how a bug occurs and how to find new bugs based on that understanding. Similarly, LLM-assisted static analysis tools~\cite{li2025iris,lekssays2025llmxcpg} use LLMs to label source and sink nodes in taint analysis, generate code queries, or triage false positives. While this reduces human effort, LLMs still serve as auxiliaries rather than to reason about why and where a bug might occur.

\PP{Insight 3: Deep agents}
Instead of limiting LLMs to assisting existing fuzzing, we use deep agents~\cite{deep-agents,claudecode,codex,gemini-cli} as the primary reasoning engine. 
Deep agent is an LLM agent paradigm that combines planning, sub-agents, filesystem access, and shell integration to execute complex tasks over a long horizon.
These capabilities are essential for finding bugs like our motivating example. Notably, such a bug cannot be discovered in a single step: the agent must analyze, hypothesize, and verify.

\subsection{Technical Challenges}
\label{ss:motiv-explore}

We combine these three insights into \emph{agentic fuzzing}, a new approach that uses deep agents to detect logic bugs through code reasoning, and implement it in \sys. Instead of receiving a test input as a seed, \sys takes \emph{a reference bug as a seed}. Given a seed, \sys analyzes its root cause, searches the codebase for other locations where the same root cause could produce a bug, and investigates each candidate to determine whether it contains a new vulnerability. Unlike traditional fuzzing, which generates or mutates inputs, and observes execution signals (e.g., crashes), \sys \emph{hypothesizes} vulnerabilities from reference bugs and \emph{verifies} them through code reasoning and PoC generation. It mimics how human auditors find bugs by analyzing, hypothesizing, and verifying.

However, implementing this approach is more challenging than it may seem, as it introduces the following challenges:

\PP{Challenge 1: Harnessing}
The first challenge is how to define tasks for \sys to perform. This is known as \emph{harness engineering}~\cite{harness-engineering}: defining the interface and behavior of an agent to perform a task. There is no algorithmic solution to this; it requires domain expertise and iterative experimentation.

We decomposed the task into a four-stage pipeline: \Analyzer, \Investigator, \ScenarioAnalyzer, and \Validator (\autoref{ss:design-pipeline}), so that each agent can focus on a single objective. As a result, our pipeline achieved the highest per-seed bug discovery (9 bugs from 100 seeds) in our evaluation (\autoref{ss:eval-pipeline}). However, the optimal harness design for agentic fuzzing remains an open problem: a single-agent baseline that self-decomposes using subagents performed comparably to our pipeline, suggesting that there may be other effective ways to structure the agents and their interactions.

\PP{Challenge 2: Repeated scenarios}
When \sys processes multiple seeds, it often produces repeated scenarios. This happens because \sys analyzes each seed independently, leading to overlapping hypotheses if the seeds are semantically similar.
This is especially common when multiple seeds report different symptoms of the same root cause (e.g., fuzzer crashes that hit the same code path) or are variants of each other.
One na\"ive way to resolve this is to append previous hypotheses to the prompt, but this increases the cost of analysis and fills the context window.

To mitigate this, \sys tracks \emph{scenario coverage}, a record of all previously explored location-hypothesis pairs.
In detail, before exploring a new hypothesis, \sys checks whether a similar hypothesis has already been investigated at that location.
If so, \sys skips the redundant investigation and instead uses a summary of prior findings to refine or devise new hypotheses.
By doing so, \sys can avoid redundant investigations and save LLM budget.

\PP{Challenge 3: Seed scheduling}
Given a large corpus of reference bugs, \sys must decide which seeds to process and in what order to maximize bug discovery within a limited budget.
Na\"ive strategies, such as processing seeds chronologically or randomly, risk concentrating effort on similar seeds, limiting the diversity of discovered bugs.
Notably, many bugs with similar root causes are clustered together in time~\cite{bugtimecluster1,bugtimecluster2,bugtimecluster3,bugtimecluster4,bugtimecluster5}, as they often originate from the same fuzzing campaign or code audit.
Ideally, we would prioritize seeds that are most likely to lead to new discoveries, but it is difficult to estimate this a priori.

Instead, we use a strategy of dispersing seed selection across different root causes to maximize diversity (\autoref{sss:seed-scheduling}). Specifically, we use the DPP-MAP~\cite{macchi1975coincidence,chen2018fast,han2020map} algorithm, which has been used in recommendation systems~\cite{chen2018fast} and machine learning~\cite{han2020map} for selecting diverse subsets of items. This approach allows \sys to cover a broad range of root causes early on, rather than spending budget on clusters of similar seeds. However, optimal seed scheduling for agentic fuzzing remains an open problem: more sophisticated strategies that estimate the potential of each seed to lead to new discoveries could further improve efficiency.

\begin{figure*}[t]
    \centering
    \includegraphics[width=\textwidth]{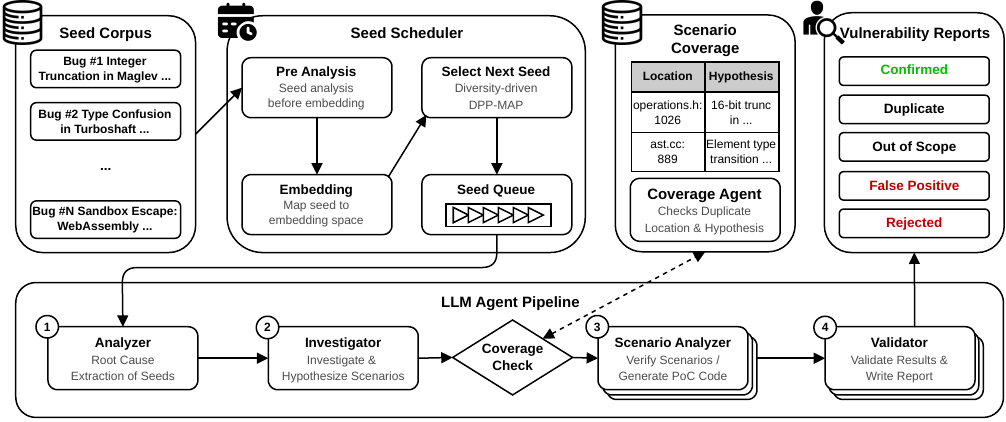}
    \caption{Overview of \sys, an Agentic Fuzzer}
    \label{fig:overview}
\end{figure*}

\subsection{Comparison to Google Big Sleep}

Recently, many works have applied LLM agents to vulnerability discovery~\cite{bigsleep2024,demoor2024xbow,deng2024pentestgpt,li2025iris,lekssays2025llmxcpg}. Among them, Google Big Sleep~\cite{bigsleep2024} is the closest work to ours, as it uses LLM agents for discovering bugs from reference bugs.
Google Big Sleep demonstrates its effectiveness by discovering vulnerabilities in multiple open-source projects, including SQLite~\cite{sqlite} and Chromium~\cite{chromium}.

Despite these similarities, we still believe our work remains novel for the following reasons.
First, \emph{Google Big Sleep} does not disclose its architecture or details. In contrast, we disclose \sys's architecture and details and leverage commercial LLMs, making it more accessible to the community.
Second, Google Big Sleep already performed vulnerability discovery on the V8 engine as shown in its issue tracker~\cite{bigsleeptracker}. In particular, the reference bug in our motivating example (CVE-2025-10892, \autoref{fig:motiv-ref}) was itself discovered by Big Sleep~\cite{bigsleeptracker}, indicating that Big Sleep has investigated this family of bugs in Maglev. However, the Turboshaft variant (\autoref{fig:motiv-vuln}), which shares the same root-cause pattern but arises in a different compiler backend, remained undiscovered until \sys found it.
Third, \sys assumes a general user setting and introduces several optimizations. Model providers (e.g., Google) face fewer constraints on token usage. In contrast, general users must manage costs, making \sys's approach more meaningful for them.

\subsection{Target Selection}
\label{ss:target}

We target the V8 JavaScript engine, a widely used open-source project with a well-maintained issue tracker~\cite{chromiumissuetracker}. As discussed above, V8's multi-tier execution pipeline creates cross-component complexity where logic bugs can hide in interactions between very different implementations. V8 is also one of the most heavily fuzzed codebases in the world: Google's ClusterFuzz~\cite{clusterfuzz} infrastructure continuously fuzzes it, and the V8 team integrates and maintains both conventional (e.g., libFuzzer~\cite{libfuzzer}) and state-of-the-art fuzzers (e.g., Fuzzilli~\cite{gross2023fuzzilli} and DUMPLING~\cite{wachter2025dumpling}). Any bug that survives this process is one that conventional fuzzing alone rarely finds. To confirm this, we ran Fuzzilli~\cite{gross2023fuzzilli} for 72 hours and it found no bugs while achieving 17.85\% code coverage. This makes V8 a good target for showing the effectiveness of agentic fuzzing in finding logic bugs that are missed by conventional fuzzing.
Although we use V8 as our primary target, \sys can generalize to other targets as we show in other JavaScript engines (\autoref{ss:eval-cross-engine}) and applications (\autoref{ss:eval-additional}).

%% file: code/motiv-vuln.cpp.tex
\begin{Verbatim}[commandchars=\\\{\},codes={\catcode`\$=3\catcode`\^=7\catcode`\_=8\relax}]
\PY{c+c1}{// src/compiler/turboshaft/operations.h}
\PY{k}{const}\PY{+w}{ }\PY{k+kt}{uint16\PYZus{}t}\PY{+w}{ }\PY{n}{input\PYZus{}count}$\tikzmark{vfld}$\PY{p}{;}\PY{+w}{ }\PY{c+c1}{// input\PYZus{}count is defined as uint16\PYZus{}t$\label{line:input-count}$}

\PY{k}{explicit}\PY{+w}{ }\PY{n}{Operation}\PY{p}{(}\PY{n}{Opcode}\PY{+w}{ }\PY{n}{opcode}\PY{p}{,}\PY{+w}{ }\PY{k+kt}{size\PYZus{}t}\PY{+w}{ }\PY{n}{input\PYZus{}count}\PY{p}{)}
\PY{+w}{    }\PY{o}{:}\PY{+w}{ }\PY{n}{opcode}\PY{p}{(}\PY{n}{opcode}\PY{p}{)}\PY{p}{,}\PY{+w}{ }\PY{n}{input\PYZus{}count}\PY{p}{(}\PY{n}{input\PYZus{}count}$\tikzmark{vinit}$\PY{p}{)}\PY{+w}{ }\PY{p}{\PYZob{}}\PY{+w}{ }\PY{c+c1}{// accepts as size\PYZus{}t$\label{line:ctor-narrow}$}
\PY{+w}{  }\PY{n}{DCHECK\PYZus{}LE}\PY{p}{(}\PY{n}{input\PYZus{}count}\PY{p}{,}\PY{+w}{ }\PY{c+c1}{// DCHECK only fires in debug builds$\label{line:dcheck}$}
\PY{+w}{            }\PY{n}{std}\PY{o}{:}\PY{o}{:}\PY{n}{numeric\PYZus{}limits}\PY{o}{\PYZlt{}}\PY{k}{decltype}\PY{p}{(}\PY{k}{this}\PY{o}{\PYZhy{}}\PY{o}{\PYZgt{}}\PY{n}{input\PYZus{}count}\PY{p}{)}\PY{o}{\PYZgt{}}\PY{o}{:}\PY{o}{:}\PY{n}{max}\PY{p}{(}\PY{p}{)}\PY{p}{)}\PY{p}{;}
\PY{p}{\PYZcb{}}
\end{Verbatim}

%% file: code/motiv-trigger.wat.tex
\begin{Verbatim}[commandchars=\\\{\},codes={\catcode`\$=3\catcode`\^=7\catcode`\_=8\relax}]
\PY{p}{(}\PY{k}{block} \PY{n+nv}{\PYZdl{}outer} \PY{p}{(}\PY{k}{result} \PY{k+kt}{i32}\PY{p}{)}
  \PY{p}{(}\PY{n+nb}{local.get} \PY{n+nv}{\PYZdl{}cond}\PY{p}{)}
  \PY{p}{(}\PY{k}{if}
    \PY{p}{(}\PY{k}{then}
      \PY{p}{(}\PY{n+nb}{i32.const} \PY{l+m+mi}{1}\PY{p}{)}
      \PY{p}{(}\PY{n+nb}{local.get} \PY{n+nv}{\PYZdl{}idx}\PY{p}{)}
      \PY{c+c1}{;; 33,000 targets, all branch to outer}
      \PY{p}{(}\PY{n+nb}{br\PYZus{}table} \PY{n+nv}{\PYZdl{}outer} \PY{n+nv}{\PYZdl{}outer} \PY{n+nv}{\PYZdl{}outer} \PY{n+nv}{\PYZdl{}outer} \PY{n}{.}\PY{n}{.}\PY{n}{.} \PY{p}{(}\PY{l+m+mf}{33,000}\PY{n}{x}\PY{p}{)}\PY{p}{)}\PY{p}{)}
    \PY{p}{(}\PY{k}{else}
      \PY{p}{(}\PY{n+nb}{i32.const} \PY{l+m+mi}{2}\PY{p}{)}
      \PY{p}{(}\PY{n+nb}{local.get} \PY{n+nv}{\PYZdl{}idx}\PY{p}{)}
      \PY{c+c1}{;; 33,000 targets, all branch to outer}
      \PY{p}{(}\PY{n+nb}{br\PYZus{}table} \PY{n+nv}{\PYZdl{}outer} \PY{n+nv}{\PYZdl{}outer} \PY{n+nv}{\PYZdl{}outer} \PY{n+nv}{\PYZdl{}outer} \PY{n}{.}\PY{n}{.}\PY{n}{.} \PY{p}{(}\PY{l+m+mf}{33,000}\PY{n}{x}\PY{p}{)}\PY{p}{)}\PY{p}{)}
  \PY{p}{)}
  \PY{p}{(}\PY{n+nb}{i32.const} \PY{l+m+mi}{0}\PY{p}{)} \PY{c+c1}{;; fallthrough}
\PY{p}{)}
\end{Verbatim}

%% file: code/motiv-ref.cpp.tex
\begin{Verbatim}[commandchars=\\\{\},codes={\catcode`\$=3\catcode`\^=7\catcode`\_=8\relax}]
\PY{c+c1}{// src/maglev/maglev\PYZhy{}ir.h}
\PY{c+c1}{// Bitfield: [opcode:16][input\PYZus{}count:16][properties:16][extras:16]}
\PY{k}{using}\PY{+w}{ }\PY{n}{OpcodeField}\PY{+w}{ }\PY{o}{=}\PY{+w}{ }\PY{n}{base}\PY{o}{:}\PY{o}{:}\PY{n}{BitField64}\PY{o}{\PYZlt{}}\PY{n}{Opcode}\PY{p}{,}\PY{+w}{ }\PY{l+m+mi}{0}\PY{p}{,}\PY{+w}{ }\PY{l+m+mi}{16}\PY{o}{\PYZgt{}}\PY{p}{;}
\PY{k}{using}\PY{+w}{ }\PY{n}{InputCountField}$\tikzmark{mfld}$\PY{+w}{ }\PY{o}{=}\PY{+w}{ }\PY{n}{OpcodeField}\PY{o}{:}\PY{o}{:}\PY{n}{Next}\PY{o}{\PYZlt{}}\PY{k+kt}{size\PYZus{}t}\PY{p}{,}\PY{+w}{ }\PY{l+m+mi}{16}\PY{o}{\PYZgt{}}\PY{p}{;}\PY{+w}{ }\PY{c+c1}{// 16\PYZhy{}bit$\label{line:maglev-bitfield}$}

\PY{c+c1}{// src/maglev/maglev\PYZhy{}ir.h (NodeBase::Allocate)}
\PY{k+kt}{uint64\PYZus{}t}\PY{+w}{ }\PY{n}{bitfield}\PY{+w}{ }\PY{o}{=}\PY{+w}{ }\PY{n}{OpcodeField}\PY{o}{:}\PY{o}{:}\PY{n}{encode}\PY{p}{(}\PY{n}{opcode\PYZus{}of}\PY{o}{\PYZlt{}}\PY{n}{Derived}\PY{o}{\PYZgt{}}\PY{p}{)}\PY{+w}{ }\PY{o}{|}
\PY{+w}{    }\PY{n}{OpPropertiesField}\PY{o}{:}\PY{o}{:}\PY{n}{encode}\PY{p}{(}\PY{n}{Derived}\PY{o}{:}\PY{o}{:}\PY{n}{kProperties}\PY{p}{)}\PY{+w}{ }\PY{o}{|}
\PY{+w}{    }\PY{n}{InputCountField}\PY{o}{:}\PY{o}{:}\PY{n}{encode}\PY{p}{(}\PY{n}{input\PYZus{}count}$\tikzmark{mic}$\PY{p}{)}\PY{p}{;}\PY{+w}{ }\PY{c+c1}{// encodes to bitfield$\label{line:maglev-encode}$}

\PY{c+c1}{// src/base/bit\PYZhy{}field.h}
\PY{k}{static}\PY{+w}{ }\PY{k}{constexpr}\PY{+w}{ }\PY{n}{U}\PY{+w}{ }\PY{n+nf}{encode}\PY{p}{(}\PY{n}{T}\PY{+w}{ }\PY{n}{value}$\tikzmark{mparam}$\PY{p}{)}\PY{+w}{ }\PY{p}{\PYZob{}}
\PY{+w}{  }\PY{n}{DCHECK}\PY{p}{(}\PY{n}{is\PYZus{}valid}\PY{p}{(}\PY{n}{value}\PY{p}{)}\PY{p}{)}\PY{p}{;}\PY{+w}{ }\PY{c+c1}{// only in debug builds$\label{line:maglev-dcheck}$}
\PY{+w}{  }\PY{k}{return}\PY{+w}{ }\PY{k}{static\PYZus{}cast}\PY{o}{\PYZlt{}}\PY{n}{U}\PY{o}{\PYZgt{}}\PY{p}{(}\PY{n}{value}\PY{p}{)}$\tikzmark{mcast}$\PY{+w}{ }\PY{o}{\PYZlt{}}\PY{o}{\PYZlt{}}\PY{+w}{ }\PY{n}{kShift}\PY{p}{;}\PY{+w}{ }\PY{c+c1}{// truncates to 16\PYZhy{}bit$\label{line:maglev-truncate}$}
\PY{p}{\PYZcb{}}

\PY{c+c1}{// src/maglev/maglev\PYZhy{}graph\PYZhy{}builder.cc}
\PY{k+kt}{int}\PY{+w}{ }\PY{n}{input\PYZus{}count}\PY{+w}{ }\PY{o}{=}\PY{+w}{ }\PY{n}{parameter\PYZus{}count\PYZus{}without\PYZus{}receiver}\PY{p}{(}\PY{p}{)}
\PY{+w}{    }\PY{o}{+}\PY{+w}{ }\PY{n}{args}\PY{p}{.}\PY{n}{register\PYZus{}count}\PY{p}{(}\PY{p}{)}
\PY{+w}{    }\PY{o}{+}\PY{+w}{ }\PY{n}{GeneratorStore}\PY{o}{:}\PY{o}{:}\PY{n}{kFixedInputCount}\PY{p}{;}\PY{+w}{ }\PY{c+c1}{// input count calculation$\label{line:maglev-compute}$}
\PY{n}{AddNewNode}\PY{o}{\PYZlt{}}\PY{n}{GeneratorStore}\PY{o}{\PYZgt{}}\PY{p}{(}\PY{n}{input\PYZus{}count}$\tikzmark{msrc}$\PY{p}{,}\PY{+w}{ }\PY{p}{.}\PY{p}{.}\PY{p}{.}\PY{p}{)}\PY{p}{;}\PY{+w}{ }\PY{c+c1}{// passed as int$\label{line:maglev-create}$}
\end{Verbatim}

%% file: code/motiv-ref-trigger.js.tex
\begin{Verbatim}[commandchars=\\\{\},codes={\catcode`\$=3\catcode`\^=7\catcode`\_=8\relax}]
\PY{k+kd}{const}\PY{+w}{ }\PY{n+nx}{kNumRegs}\PY{+w}{ }\PY{o}{=}\PY{+w}{ }\PY{l+m+mf}{65534}\PY{p}{;}
\PY{k+kd}{let}\PY{+w}{ }\PY{n+nx}{body}\PY{+w}{ }\PY{o}{=}\PY{+w}{ }\PY{p}{[}\PY{p}{]}\PY{p}{;}
\PY{k}{for}\PY{+w}{ }\PY{p}{(}\PY{k+kd}{let}\PY{+w}{ }\PY{n+nx}{i}\PY{+w}{ }\PY{o}{=}\PY{+w}{ }\PY{l+m+mf}{0}\PY{p}{;}\PY{+w}{ }\PY{n+nx}{i}\PY{+w}{ }\PY{o}{\PYZlt{}}\PY{+w}{ }\PY{n+nx}{kNumRegs}\PY{p}{;}\PY{+w}{ }\PY{o}{++}\PY{n+nx}{i}\PY{p}{)}\PY{+w}{ }\PY{p}{\PYZob{}}
\PY{+w}{  }\PY{n+nx}{body}\PY{p}{.}\PY{n+nx}{push}\PY{p}{(}\PY{l+s+sb}{`}\PY{l+s+sb}{  let r}\PY{l+s+si}{\PYZdl{}\PYZob{}}\PY{n+nx}{i}\PY{l+s+si}{\PYZcb{}}\PY{l+s+sb}{ = }\PY{l+s+si}{\PYZdl{}\PYZob{}}\PY{n+nx}{i}\PY{l+s+si}{\PYZcb{}}\PY{l+s+sb}{;}\PY{l+s+sb}{`}\PY{p}{)}\PY{p}{;}
\PY{p}{\PYZcb{}}
\PY{k+kd}{let}\PY{+w}{ }\PY{n+nx}{f}\PY{+w}{ }\PY{o}{=}\PY{+w}{ }\PY{n+nb}{eval}\PY{p}{(}\PY{l+s+sb}{`}\PY{l+s+sb}{(function*() \PYZob{}}\PY{l+s+sb}{\PYZbs{}n}\PY{l+s+si}{\PYZdl{}\PYZob{}}\PY{n+nx}{body}\PY{p}{.}\PY{n+nx}{join}\PY{p}{(}\PY{l+s+s1}{\PYZsq{}\PYZbs{}n\PYZsq{}}\PY{p}{)}\PY{l+s+si}{\PYZcb{}}\PY{l+s+sb}{\PYZcb{})}\PY{l+s+sb}{`}\PY{p}{)}\PY{p}{;}

\PY{o}{\PYZpc{}}\PY{n+nx}{PrepareFunctionForOptimization}\PY{p}{(}\PY{n+nx}{f}\PY{p}{)}\PY{p}{;}
\PY{n+nx}{f}\PY{p}{(}\PY{p}{)}\PY{p}{;}
\PY{o}{\PYZpc{}}\PY{n+nx}{OptimizeMaglevOnNextCall}\PY{p}{(}\PY{n+nx}{f}\PY{p}{)}\PY{p}{;}
\PY{n+nx}{f}\PY{p}{(}\PY{p}{)}\PY{p}{;}
\end{Verbatim}

%% file: design.tex
\section{Design}
\label{s:design}

\subsection{Overview}
\label{ss:design-overview}

\autoref{fig:overview} shows the overall architecture of \sys. \sys consists of three main components: \WC{1} the seed scheduler, \WC{2} the LLM agent pipeline, and \WC{3} the scenario coverage tracker. Given a corpus of reference bugs as seeds, \sys works as follows. Initially, the seed scheduler (\autoref{ss:design-exploration}) schedules the seeds by iteratively selecting a seed that maximizes the diversity of explored bugs. Each seed is then processed by the four-stage LLM agent pipeline (\autoref{ss:design-pipeline}), which searches for vulnerabilities in the target codebase that share similar root causes with the reference bug. Throughout this process, the scenario coverage tracker records which (location, hypothesis) pairs have been explored, allowing the pipeline to skip redundant investigations across different seeds.

\subsection{LLM Agent Pipeline}
\label{ss:design-pipeline}

The LLM agent pipeline consists of four specialized agents that execute sequentially: \WC{1} \Analyzer, \WC{2} \Investigator, \WC{3} \ScenarioAnalyzer, and \WC{4} \Validator. Each agent feeds its results to the next stage, progressively deepening the investigation from understanding the reference bug to validating new vulnerabilities. The agents operate on the target codebase, reading source files, hypothesizing scenarios, and executing test code to validate those scenarios. We describe each stage in detail below. We also explain an example of how the pipeline discovers a bug in \autoref{s:appendix:design_example}.

\PP{Stage 1: Analyzer}
\Analyzer takes a reference bug as input and produces a comprehensive analysis report. Given a reference bug as an issue tracker URL or a commit hash, \Analyzer fetches related artifacts such as the bug report, patches, code reviews, and PoC code. Then, it examines the relevant source code in the target codebase to understand the vulnerability. As a result, \Analyzer produces a report containing comprehensive information about the bug, including the root cause, bug mechanism, impact, fix description, affected files and functions, vulnerable code snippets, and patch code. One notable output is the \emph{bug mechanism}, which describes the step-by-step process of how the vulnerability works. This mechanism guides the agents in later stages to search for similar instances of the bug.

\PP{Stage 2: Investigator}
\Investigator takes the analysis report from \Analyzer and searches the target codebase for locations where the same or similar assumption violations might exist. Guided by the bug mechanism, \Investigator reads source files, traces code paths, and identifies locations that share the same underlying pattern as the reference bug. Specifically, it looks for direct pattern matches, subclasses or implementations of the same vulnerable interfaces, related subsystems with similar logic, and cross-component interaction boundaries. When \Investigator finds a suspicious location, it formulates a vulnerability hypothesis describing the broken assumption or missing validation and submits it as a \emph{scenario} for the next stage. Each scenario includes the affected source locations, a potential trigger path from JavaScript to the vulnerable code, and advisory notes that help \ScenarioAnalyzer navigate defensive checks or target specific optimization states. \Investigator does not stop at the first scenario; it keeps exploring the codebase for multiple independent scenarios before completing its investigation.

\PP{Stage 3: Scenario Analyzer}
\ScenarioAnalyzer receives a scenario from \Investigator and verifies whether the hypothesis in the scenario is valid by producing a Proof-of-Concept (PoC) that demonstrates the vulnerability. It first reads the scenario's code locations in the target codebase, traces the execution path from JavaScript to the suspected vulnerable code, and determines what conditions are needed to reach it. Using the examined trigger path and advisory notes from \Investigator as a starting point, \ScenarioAnalyzer writes an initial batch of test code. It then executes the test code on the target engine, checking if the execution produces observable evidence of a vulnerability such as crashes, debug assertion failures, or unexpected values that violate the JavaScript specification. If the initial attempt does not produce observable evidence, \ScenarioAnalyzer iteratively refines its approach---trying different JavaScript constructs, optimization states (interpreted, baseline-compiled, or optimized code), object shapes, and edge cases. Finally, \ScenarioAnalyzer examines the execution results and reports either success, with the final PoC code and execution results, or failure, with a description of the approaches tried and why they did not work. Scenarios with a successful PoC are forwarded to \Validator for final verification.

\PP{Stage 4: Validator}
\Validator independently verifies each scenario and produces a vulnerability report for confirmed vulnerabilities. For each scenario with a successful PoC, \Validator re-executes the PoC, checks the validity of the evidence, and determines whether to accept or reject the scenario. It also checks whether the evidence conforms to the threat model (see \autoref{s:impl}). For example, it warns if the execution of the PoC uses any security-disabling flags (e.g., \cc{--no-wasm-bounds-checks}) or debug-only native functions (e.g., \cc{\%AbortJS}). For confirmed vulnerabilities, \Validator produces a vulnerability report containing summary, technical details, trigger conditions, reproduction steps with outputs from both release and debug builds, and a suggested patch. Humans can then independently validate the report and decide whether it is worth reporting to the vendor.

\subsection{Exploration Strategy}
\label{ss:design-exploration}

As discussed in Challenges 2 and 3 of \autoref{ss:motiv-explore}, scaling agentic fuzzing to a large number of seeds (i.e., reference bugs) raises two challenges: avoiding investigation of redundant scenarios and scheduling seeds. We describe our approaches to each below.

\begin{figure}[t]
    \centering
    \includegraphics[width=\columnwidth]{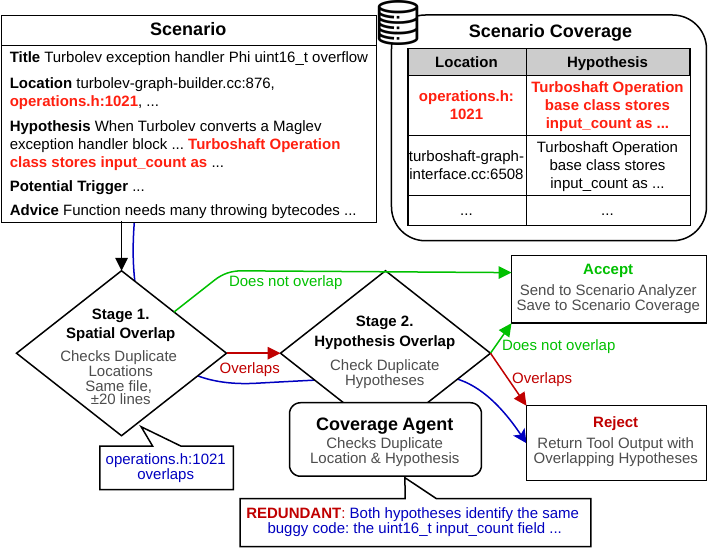}
    \caption{Scenario coverage to avoid redundant investigations. The example scenario (\autoref{ss:motiv-variant}) follows the blue arrow (Stage 1 $\to$ Stage 2 $\to$ Reject), rejected due to spatial and hypothesis overlap with a previously investigated scenario.}
    \label{fig:coverage}
\end{figure}

\subsubsection{Scenario coverage}
\label{sss:scenario-coverage}

Scenario coverage tracks all previously explored scenarios (i.e., location-hypothesis pairs) across the entire fuzzing campaign, preventing redundant investigations across different seeds. \autoref{fig:coverage} depicts how it works. Each time \Investigator submits a new scenario, the coverage tracker checks whether a similar scenario has already been explored at overlapping code locations before dispatching it to \ScenarioAnalyzer.

The check proceeds in two stages. First, it queries the coverage database for existing entries matching the source file locations of the scenario. Agents often investigate the same code region with slightly different line numbers, especially when a statement spans multiple lines. To account for this, the check considers two scenarios to be spatially overlapping if they target the same source file and their line numbers fall within a tolerance window ($\pm20$ lines). If no overlapping entry exists, the scenario is approved immediately. If overlapping entries are found, an LLM agent using a lightweight model (e.g., Haiku) compares the proposed hypothesis against existing hypotheses to determine whether they overlap semantically. Redundant scenarios are rejected; distinct ones are approved. Approved scenarios are recorded in the coverage database so that future submissions can be checked against them.

\PP{Example}
In the example of \autoref{fig:coverage}, the proposed scenario tries to achieve integer truncation using the exception handler of a different compiler component, Turbolev. It flows through Stage 1 $\to$ Stage 2 $\to$ Reject (the blue arrow in the figure), and is rejected as redundant with the existing scenario of our motivating example (\autoref{ss:motiv-variant}). At Stage 1, the two scenarios are considered spatially overlapping because they both target \cc{operations.h\:1021}. At Stage 2, the proposed scenario is considered semantically redundant with the existing scenario. Although they use different attack vectors, they share the same root cause: both hypothesize that the Turboshaft Operation class stores \cc{input\_count} as \cc{uint16\_t}, which can lead to an integer truncation vulnerability. Since any scenario that exploits this root cause would yield the same bug regardless of the attack vector, the proposed scenario is rejected and discarded.

\begin{algorithm}[t]
\caption{Seed scheduling via Fast Greedy DPP-MAP~\cite{chen2018fast}.}
\label{alg:seed-scheduling}
\algtext*{EndWhile}%
\algtext*{EndFor}%
\algtext*{EndIf}%
\begin{algorithmic}[1]
\Require Seed corpus $\mathcal{C}$, embedding model $\phi$
\Statex \textit{// Pre-analysis}
\For{each seed $s \in \mathcal{C}$}
    \State $t_s \gets \textsc{PreAnalyze}(s)$ \Comment{Pre-analyze seed (\Analyzer)}
    \State $\mathbf{v}_s \gets \phi(t_s) / \|\phi(t_s)\|$ \Comment{Normalized embedding}
\EndFor
\Statex \textit{// Iterative scheduling}
\State $\mathcal{P} \gets \emptyset$ \Comment{Already-processed seeds}
\While{$\mathcal{C} \setminus \mathcal{P} \neq \emptyset$}
    \State $s^* \gets \textsc{FastGreedyDPP}(\{\mathbf{v}_s\}_{s \in \mathcal{C} \setminus \mathcal{P}}, \{\mathbf{v}_s\}_{s \in \mathcal{P}})$ \Comment{\cite{chen2018fast}}
    \State Dispatch $s^*$ to the agent pipeline
    \State $\mathcal{P} \gets \mathcal{P} \cup \{s^*\}$
\EndWhile
\end{algorithmic}
\end{algorithm}

\subsubsection{Seed scheduling}
\label{sss:seed-scheduling}

To schedule seeds from a large corpus of reference bugs, \sys prioritizes diversity by selecting seeds that cover distinct root causes and bug mechanisms before similar ones. To achieve this, \sys uses Determinantal Point Process Maximum a Posteriori inference (DPP-MAP)~\cite{macchi1975coincidence,chen2018fast,han2020map}, which was previously applied in recommendation systems~\cite{chen2018fast} and machine learning~\cite{han2020map} for selecting diverse subsets of items.

The scheduling works in three steps as shown in Algorithm~\ref{alg:seed-scheduling}. First, \Analyzer pre-analyzes all seeds in the corpus to extract their root causes and bug mechanisms. Although pre-analyzing all seeds in advance is costly, this is a one-time cost that can be amortized across multiple runs, and the results can be directly reused as \Analyzer results during actual runs. Second, the pre-analysis results are mapped to normalized vector embeddings using a text embedding model (e.g., OpenAI's \cc{text-embedding-3-large}), so that seeds with similar root causes and mechanisms can be mapped close together in the embedding space. Finally, \sys uses the Fast Greedy DPP-MAP algorithm of Chen \etal~\cite{chen2018fast} to greedily select the seed most distinct from those already processed in the embedding space. This algorithm selects the next item that maximizes the diversity of the selected subset at each step, measured by the determinant of the kernel matrix of the selected items.

%% file: impl.tex
\section{Implementation}
\label{s:impl}

We implemented \sys as a web application with a total of 28.0k lines of code, including 13.0k lines of Python for the backend, 3.4k lines of Python and 720 lines of prompt templates for the agent service, and 10.9k lines of TypeScript for the frontend. The agents are built on the Claude Agent SDK~\cite{claudeagentsdk}, which provides an interface to Claude Code~\cite{claudecode}. The backend is responsible for agent task orchestration, database management, and API handling while the frontend provides a user interface for monitoring and interacting with running agents. We designed the backend to spawn agents as individual Docker containers, allowing for scalable and isolated execution. Each agent container has access to its own target source code and prebuilt binaries, enabling it to perform dynamic analysis and PoC execution as needed.

\PP{Time budget and soft timeout}
To prevent agents from running too long and exhausting resources, we implemented a soft timeout mechanism. Specifically, we observed that agents occasionally fall into a compaction death spiral~\cite{konstantin2025stopthebleed, claudecodeissue24179}. In this state, agents loop over the same tasks without progress as long conversations lose context during compaction. To mitigate this, \sys injects a warning into the agent's conversation at 50\%, 80\%, and 90\% of its soft time budget, telling the agent how much time remains and prompting it to wrap up. After the soft timeout, additional warnings fire every 5 minutes until the hard timeout kills the process. \sys currently sets the soft timeout to 6 hours and the hard timeout to 12 hours, but these values can be adjusted when needed.

\PP{PoC execution and threat model warning}
\sys targets bugs reachable by untrusted JavaScript, so all PoC code must trigger the vulnerability without security-disabling flags or debug-only native functions. Agents execute PoC code through a dedicated tool that runs it against prebuilt \cc{d8} binaries (the V8 developer shell). We precompiled both release and debug builds for nine architectures: x64, arm64, ia32, arm, loong64, mips64el, ppc64, s390x, and riscv64. The flags \cc{--allow-natives-syntax} and \cc{--expose-gc} are always included so that agents can use optimization-triggering natives and explicit garbage collection. We also apply a 300-second timeout to each execution to prevent hanging.

While we instruct agents to use PoC code that works without security-disabling flags or debug-only natives, our execution tool also warns about violations programmatically. Our execution tool implements this using a blocklist of flags and an allowlist of native functions. If a violation is found, the tool appends a warning to the execution result so the agent can self-correct. The flag blocklist includes \cc{--expose-memory-corruption-api}, \cc{--no-sandbox}, and any flag whose name contains ``experimental,'' since experimental features are not shipped in production. The native allowlist conservatively permits only 15 optimization-related functions (e.g., \cc{\%OptimizeFunctionOnNextCall}); all others are rejected. Beyond these validations, the tool inspects execution results for common false positives. A \cc{CHECK} failure (release-build assertion) means V8 intentionally detected the issue and terminated, which is usually not a bug; the tool warns the agent to look for a \cc{DCHECK} failure (debug-only, compiled out in release) or a crash instead. For sandbox bugs, the tool checks if the output contains ``V8 sandbox violation detected''; we ignore other sandbox-related messages and warn the agent that they indicate the sandbox is working correctly rather than being bypassed (e.g., ``Safely terminating process'').

\PP{Mitigating commit history bias}
We observed that agents tend to pipe \cc{git log} through \cc{head} to keep context short, which means they only see the most recent commits and miss older ones that may be relevant. To mitigate this, we wrapped \cc{git log} with a script that randomly shuffles the output entries before returning them. To notify agents of this behavior, the wrapper prepends a note to the output indicating that the commit history is not in chronological order. If agents need to see the commit history in its original order, they can pass a flag (i.e., \cc{--no-shuffle}).

\begin{table*}[t]
    \centering
    \footnotesize
    \caption{New bugs discovered by \sys. Ref. ID refers to the seed that led to the discovery of the bug.}
    \label{tbl:bugs}
    \input{tbl/bugs}
\end{table*}

\PP{Collecting and processing seeds}
We implemented a seed collector that fetches V8 security bugs from the Chromium issue tracker~\cite{chromiumissuetracker}. We collected \seedtotal seeds by filtering for vulnerabilities in V8 while excluding duplicate, intended-behavior, infeasible, obsolete, and non-reproducible reports. Each seed is represented by its issue URL, which \Analyzer fetches at runtime to retrieve its contents. To aid \Analyzer, we also implemented a tool that extracts the contents of a bug report from its URL, including the discussion, patches, and PoC code if available. We needed the tool because the issue tracker dynamically loads its contents, which the built-in WebFetch tool of the Claude Agent SDK~\cite{claudeagentsdk} cannot fully retrieve. For seed scheduling, we embedded the pre-analysis text of each seed using OpenAI's \cc{text-embedding-3-large} model and ran DPP-MAP over these embeddings as described in \autoref{ss:design-exploration}.

%% file: tbl/bugs.tex
\begin{tabular}{lllll}
    \toprule
    \textbf{\#} & \textbf{ID} & \textbf{Detail} & \textbf{Ref. ID} & \textbf{Status} \\
    \midrule
    1 & \bidI & Type Confusion in MegaDOM IC & 450328966 & Fixed \\
    2$\dagger$$\ast$ & \bidII & Integer Truncation in Turboshaft PhiOp input\_count via WASM br\_table & 444048019 & Fixed \\
    3 & \bidIII & TOCTOU Race During [REDACTED] & 427600180 & Duplicate \\
    4$\dagger$ & \bidIV &  Missing Guard in [REDACTED] & 458090625 & Fixed \\
    5 & \bidV & Sandbox Escape via [REDACTED] & 458679941 & Duplicate \\
    6$\diamond$ & \bidVI & Signed Integer Overflow in [REDACTED] & 444048019 & Fixed \\
    7 & \bidVII & JIT Miscompilation via [REDACTED] & 40088942 & Fixed \\
    8$\ddagger$ & \bidVIII & Incorrect [REDACTED] & 40055069 & Fixed \\
    9 & \bidIX & Signed Integer Overflow in parseInt via InternalStringToIntDouble & 420697404 & Reported \\
    10 & \bidX & DCHECK Failure in Scope::ForceDynamicLookup via PreParser info\_id\_ Reset in REPL Mode & 418478214 & Fixed \\
    11 & \bidXI & Use Count Corruption in Maglev Phi Representation Selector via HoleyFloat64 ToBoolean Path & 428226995 & Duplicate \\
    12 & \bidXII & JIT Miscompilation in Turboshaft Loop Unrolling via Non-Commutative Sub & 436305802 & Fixed \\
    13 & \bidXIII & Incomplete Constant Pool State Cleanup in ARM Assembler via Maglev Codegen Abort & 40057489 & Fixed \\
    14 & \bidXIV & DCHECK Mismatch in DependOnContextCell via Function Context Cells & 412756062 & Fixed \\
    15 & \bidXV & DCHECK Failure in Label Destructor via Regexp Compilation Error Propagation & 452681948 & Fixed \\
    16 & \bidXVI & Miscompilation in Turboshaft via Div-to-Mul with Denormal Reciprocal & 429761781 & Fixed \\
    17 & \bidXVII & Missing [REDACTED] During Deopt. & 424627229 & Fixed \\
    18 & \bidXVIII & OOB Vector Access in DataDrop During Streaming WASM Compilation & 40094133 & Fixed \\
    19 & \bidXIX & Integer Overflow in [REDACTED] & 369685641 & Fixed \\
    20 & \bidXX & Missing Prototype Chain Dependency for Module Exports via PropertyAccessInfo::ModuleExport & 40057622 & Fixed \\
    21 & \bidXXI & Spec Violation in Parser via Missing Escape Check for 'using' Contextual Keyword & 40091892 & Reported \\
    22 & \bidXXII & Multiple Disposal of Resources in C-style For Loops with \cc{using} Declarations & 385170388 & Fixed \\
    23 & \bidXXIII & Missing Immutability Check for Exported \cc{using}/\cc{await using} Module Variables & 385170388 & Fixed \\
    24 & \bidXXIV & Spec Violation in Parser via Missing PrecededBy Member Checks & 40092882 & Fixed \\
    25 & \bidXXV & Missing StackOverflowCheck in JSBuiltinsConstructStubHelper on RISCV/LOONG64/MIPS64 & 40094550 & Fixed \\
    26 & \bidXXVI & Signed/Unsigned Confusion in Fast API kUint64 Argument Lowering via CheckedSigned64AsWord64 & 40064983 & Fixed \\
    27 & \bidXXVII & Incorrect Private Name Resolution in Heritage Expression via Skip Bit Loss in FinalizeBlockScope & 40093214 & Fixed \\
    28 & \bidXXVIII & Incorrect Deoptimization of \cc{undefined} to \cc{NaN} via Missing Check in Constant Encoding & 394120836 & Reported \\
    29 & \bidXXIX & Incorrect Math.round in Maglev on RISCV64/LOONG64 via Flawed "add 0.5 then floor" Algorithm & 40063144 & Fixed \\
    30 & \bidXXX & Incorrect ToUint8Clamped Codegen in RISCV64 Maglev Backend & 380604249 & Fixed \\
    31 & \bidXXXI & Incorrect Codegen in RISCV64 Maglev \cc{CompareIntPtrAndBranch} Uses Uninitialized Scratch Register & 380604249 & Reported \\
    32 & \bidXXXII & Spec Violation in IterableForEach --- Missing Prototype Check Bypasses Iterator Protocol & 385386138 & Fixed \\
    33 & \bidXXXIII & Register Clobber in RISCV UnalignedStoreHelper via Hardcoded t4 Scratch Register & 380604249 & Reported \\
    34 & \bidXXXIV & Spec Violation in Float16Array Search Methods via Incorrect Range Check & 394120836 & Fixed \\
    35$\S$ & \bidXXXV & Race Condition in WASM Stack Switching via Profiler Signal Handler & 424905890 & Confirmed \\
    36$\S$ & \bidXXXVI & Type Confusion in [REDACTED] & 420464880 & Confirmed \\
    37$\S$ & \bidXXXVII & Incorrect F16x8 Comparison in TurboFan/Liftoff via Incorrect vpackssdw Usage & 369685641 & Reported \\
    38$\S$ & \bidXXXVIII & ImmutableArrayBuffer Write Protection Bypass in WASM Turboshaft DataView Setter & 384773802 & Fixed \\
    39$\S$ & \bidXXXIX & Register Allocation Mismatch in F16x8ReplaceLane Instruction Selector on x64 & 40052865 & Reported \\
    40$\S$ & \bidXL & SIGSEGV/DCHECK Failure in SetOrCopyDataProperties via Object.assign on Shared-Space Objects & 40062686 & Fixed \\
    \bottomrule
    \multicolumn{5}{l}{
    $\ast$ Assigned \cveII.
    $\dagger$ Each awarded \$11,000 bounty.
    $\ddagger$ Awarded \$8,000 bounty and \cveVIII.
    $\diamond$ Awarded \$5,000 bounty.
    $\S$ Requires experimental flag.
    }
\end{tabular}

%% file: eval.tex
\section{Evaluation}
\label{s:eval}

\PP{Research questions}
In this section, we answer the research questions below.
\begin{itemize}[leftmargin=*]
  \item \textbf{RQ1:} Can \sys find new bugs in V8? (\autoref{ss:eval-newbugs})
  \item \textbf{RQ2:} How do pipeline design choices affect bug-finding effectiveness? (\autoref{ss:eval-pipeline})
  \item \textbf{RQ3:} Does DPP-MAP seed scheduling improve exploration diversity over simpler strategies? (\autoref{ss:eval-seed-scheduling})
  \item \textbf{RQ4:} How much do \Analyzer stage and the reference bugs contribute to bug discovery? (\autoref{ss:eval-reference-bugs})
  \item \textbf{RQ5:} Can seeds from one engine be used to find bugs in other JavaScript engines? (\autoref{ss:eval-cross-engine})
\end{itemize}
We also include our mini-experiment results with open-source models and other software targets in \autoref{ss:eval-additional}.

\subsection{Experimental Setup}

\PP{Environments}
We conducted all our experiments on servers running Ubuntu 24.04.3, each equipped with two Intel Xeon Gold 6248R CPUs (48 cores total) and 256 GB of RAM. As described in \autoref{s:impl}, agents run in isolated Docker containers limited to 6 hours of soft time and 12 hours of hard time. We limited each agent to 32 GB of RAM to prevent resource exhaustion. For LLM API calls, we used Claude Opus 4.6 via the Claude Agent SDK~\cite{claudeagentsdk}, which provides an interface to Claude Code~\cite{claudecode}. We used OpenAI's \cc{text-embedding-3-large} model for text embeddings in seed scheduling. We evaluated on the V8 JavaScript engine (\autoref{ss:target}), with additional experiments on SpiderMonkey and JavaScriptCore (\autoref{ss:eval-cross-engine}).

\PP{Seed corpus and pre-analysis}
We collected a total of \seedtotal seeds from the Chromium issue tracker (see \autoref{s:impl}). For RQ1 (\autoref{ss:eval-newbugs}), we used the full corpus of \seedtotal seeds to maximize the chance of finding new bugs in V8, though we managed to run about 750 of them within our budget. For RQ2--RQ5, we could not afford the pre-analysis cost of seed scheduling, which requires running \Analyzer on each seed to extract its root cause and bug mechanism (see \autoref{ss:design-exploration}). Therefore, we limited the scope of our evaluation to \seedeval seeds selected from the main components of V8: parser, interpreter, and compilers.

For RQ2--RQ5, we pre-analyzed \seedeval seeds using the Claude Opus 4.6 model with high effort, which cost \$1,068 in API calls (\$1.60 per seed on average) and took 5,526 minutes (8.3 minutes per seed on average). We then embedded the pre-analysis outputs using OpenAI's \cc{text-embedding-3-large} model at negligible cost (under \$0.0001) and ran DPP-MAP over the resulting embeddings. Note that pre-analysis is a one-time cost per seed: once a seed is pre-analyzed and embedded, its embedding can be reused across multiple scheduling rounds without additional API calls.

\subsection{Threat to Validity}
\label{ss:eval-threats}
\PP{Budget constraints}
Our budget limited the accuracy, breadth, and depth of our evaluation. We had to restrict our evaluation to the \seedeval seeds and the Claude Opus 4.6 model, and capped most experiments at \$1{,}000 or 100 seeds. Also, because LLM outputs are non-deterministic, a larger number of repetitions would be needed for statistical confidence, but our budget did not allow this. A larger budget would also have allowed us to evaluate on more seeds and compare across different models.

\PP{Tool and model updates}
Agent behavior can change due to updates to the tools and models. We observed that updates to Claude Code (e.g., the addition of general-purpose subagents~\cite{claudecodesubagents}) can alter our agents' behavior. Therefore, for evaluation, we pinned the versions of Claude Agent SDK~\cite{claudeagentsdk} and Claude Code~\cite{claudecode} to v0.1.49 and v2.1.77, respectively, which were the latest versions available at the time. However, our results are still susceptible to silent model changes that may have occurred during our evaluation, as some public reports claim that model behavior changed~\cite{claudecodeissue42796}.

\subsection{RQ1. Discovered New Bugs by \sys}
\label{ss:eval-newbugs}

Over a three-month span from 2026-02 to 2026-04, we ran \sys for about one month on about 750 of the \seedtotal seeds collected from the Chromium issue tracker~\cite{chromiumissuetracker}. During this period, \sys discovered \bugtotal bugs as listed in \autoref{tbl:bugs}, including \bugduplicate duplicates.
So far, \bugfixed have been fixed, \bugconfirmed confirmed, and the remaining \bugreview non-duplicate bugs are still under review. Four of the bugs received VRP bounties: bug \#2 (\$11{,}000), bug \#4 (\$11{,}000), bug \#8 (\$8{,}000), and bug \#6 (\$5{,}000), totaling \$35{,}000. Bug \#2 and \#8 were each assigned \cveII and \cveVIII respectively. Although we instructed agents to avoid using experimental flags as explained in \autoref{s:impl}, they occasionally ignored our instructions during exploration and found \bugexperimental bugs that require an experimental flag to trigger. We still reported these bugs as they were valid and should be fixed before the features are enabled by default. These results show that \sys can find new bugs in V8, including security vulnerabilities that were not previously reported.

\sys reported bugs in various V8 subsystems, including the optimizing compilers (e.g., Maglev and TurboFan), the parser, the interpreter, the WASM pipeline, and architecture-specific backends. The bugs also cover various types of issues, from specification violations to exploitable type confusions. We provide detailed case studies of two interesting bugs in \autoref{s:appendix:case-studies}.

\begin{table}[t]
    \centering
    \footnotesize
    \caption{Pipeline design choice comparison. \cc{\#} is the number of seeds processed. \cc{TP} and \cc{FP} are true and false positives.}
    \label{tbl:pipeline}
    \input{tbl/pipeline}
\end{table}

\begin{table}[t]
    \centering
    \footnotesize
    \caption{Bug reproducibility across pipeline design choices. \cc{pass@5} counts how many of the 21 bugs were rediscovered at least once across five runs; \cc{Total} counts the total rediscoveries across all runs. For per-bug detail, see \autoref{tbl:reproducibility-full} of \appendixref.}
    \label{tbl:reproducibility}
    \input{tbl/reproducibility}
\end{table}

\subsection{RQ2. Pipeline Design Choices}
\label{ss:eval-pipeline}

\PP{Experiment setup}
\autoref{tbl:pipeline} compares three pipeline designs using Claude Opus 4.6 at three effort levels (high, medium, and low), which control the model's reasoning depth, with our DPP-MAP scheduling. The four-stage pipeline is the full \sys pipeline (\Analyzer $\to$ \Investigator $\to$ \ScenarioAnalyzer $\to$ \Validator) as described in \autoref{ss:design-pipeline}. The three-stage pipeline (\Analyzer $\to$ \BugFinder $\to$ \Validator) merges \Investigator and \ScenarioAnalyzer into a single \BugFinder agent that both explores the codebase and generates PoCs. The one-stage pipeline runs a single \BugFinder agent that performs analysis, investigation, PoC generation, and validation within one execution. For fair comparison, we included all detailed prompts and instructions of \sys for \BugFinder agent in the three- and one-stage designs, except for the explicit stage descriptions. This way, they can have similar information and guidance to the four-stage pipeline. However, because \BugFinder agent does not explicitly formulate scenarios, the scenario coverage (\autoref{ss:design-exploration}) is not used in the three- and one-stage designs.

We evaluated each design in two ways. First, we ran each design once under two resource settings (\autoref{tbl:pipeline}): (a) a fixed API budget of \$1{,}000, and (b) a fixed count of 100 seeds. Second, because LLM-based agents are non-deterministic, we additionally measured the reproducibility of discovered bugs across pipeline designs and effort levels. We took the seeds that produced bug discoveries in the first experiment (18 mapped to \autoref{tbl:bugs}; 3 were independently fixed by V8 developers before we reported them), reran them five times each, and counted how often each bug was rediscovered (\autoref{tbl:reproducibility}; see \autoref{tbl:reproducibility-full} of \appendixref for detail).

\PP{Results}
Our results show that no single pipeline design dominates across both resource settings. The depth of the pipeline trades off against the number of seeds a fixed budget can process. Under a fixed seed count of 100, the four-stage pipeline at high effort found the most bugs (9, vs.\ 8 for one-stage and 3 for three-stage) and achieved the highest rediscovery rates across five runs (15 of 21 at pass@5, 32 total discoveries out of 105). However, it spent \$2{,}846, roughly $2.4\times$ the one-stage pipeline's \$1{,}189. Under a fixed budget of \$1{,}000, this cost difference reverses the ranking: the one-stage pipeline at high effort found 8 bugs from 82 seeds while the four-stage pipeline found only 3 from 39 seeds. We observed that the one-stage agent frequently invokes auxiliaries like subagents~\cite{claudecodesubagents} and todo list management~\cite{claudeagentsdktodo}, which let a single agent decompose the task and manage its workflow without explicit stage separation. In short, the best pipeline design depends on whether the bottleneck is seeds (prefer four-stage, more bugs per seed) or dollars (prefer one-stage, more bugs per dollar).

\subsection{RQ3. Seed Scheduling Strategies}
\label{ss:eval-seed-scheduling}

\PP{Experiment setup}
To evaluate the effectiveness of our DPP-MAP seed scheduling strategy (\autoref{ss:design-exploration}), we compared it against two other strategies: Newest-first and Random. Newest-first prioritizes seeds based on their creation date, with newer seeds being processed first. Random selects seeds randomly from the corpus without any specific ordering. We ran each strategy under the same two resource settings as in \autoref{ss:eval-pipeline}: (a) a fixed API budget of \$1,000, and (b) a fixed count of 100 seeds. All runs used Claude Opus 4.6 with the medium effort level, which is the default.

\begin{table}[t]
    \centering
    \footnotesize
    \caption{Evaluation of seed scheduling strategies. \cc{\#} is the number of seeds in each run. \cc{Cov.} measures the fraction of scenarios that pass the coverage redundancy check.}
    \label{tbl:seed-scheduling}
    \resizebox{\columnwidth}{!}{
        \input{tbl/seed-scheduling}
    }
\end{table}

\PP{Results}
Among the three strategies, DPP-MAP shows the highest scenario approval rates (99.0\%$^{(\text{a})}$ for the fixed API budget and 94.4\%$^{(\text{b})}$ for the fixed seed count), compared to Newest-first (83.2\%$^{(\text{a})}$/80.0\%$^{(\text{b})}$) and Random (90.2\%$^{(\text{a})}$/90.7\%$^{(\text{b})}$). The higher approval rates of DPP-MAP indicate that its selected seeds produce scenarios that overlap less with previously explored ones, confirming that our diversity-based scheduling reduces redundant investigations. This was also reflected in the average cost and time per seed: DPP-MAP is the most efficient with \$18.8$^{(\text{a})}$/\$19.1$^{(\text{b})}$ and 80.5$^{(\text{a})}$/84.8$^{(\text{b})}$ minutes, while the other two strategies cost more (\$22.4$^{(\text{a})}$/\$22.6$^{(\text{b})}$ and \$19.5$^{(\text{a})}$/\$19.4$^{(\text{b})}$) and take longer (96.0$^{(\text{a})}$/102.8$^{(\text{b})}$ and 89.1$^{(\text{a})}$/92.4$^{(\text{b})}$ minutes). Unfortunately, even though DPP-MAP showed better efficiency, it did not find more bugs than the other strategies in our experiments. We attribute this to the small absolute bug counts (one to six per strategy), which make the differences inconclusive. Due to the budget limit described in \autoref{ss:eval-threats}, we could not run more seeds or repeat runs to achieve statistical confidence.

\subsection{RQ4. Effect of Reference Bugs}
\label{ss:eval-reference-bugs}

\PP{Experiment setup}
To measure how much reference bugs contribute to bug discovery, we disabled \Analyzer stage and ran the remaining three stages. We instructed \Investigator to explore the parser, interpreter, and compiler subsystems of V8 for bugs, following the same scope as our seed corpus. We ran experiments with Claude Opus 4.6 at three effort levels (high, medium, and low) under a fixed API budget of \$1,000. At the high and medium efforts, we included one additional iteration that exceeded the budget because excluding it would leave the budget significantly unused.

\PP{Results}
\autoref{tbl:reference-bugs} shows that \Analyzer stage and reference bugs contribute to bug discovery. Removing them resulted in discovering fewer bugs at each effort level: 1, 1, and 0 bugs at high, medium, and low effort, respectively, compared to 3, 2, and 4 bugs in \sys. Without reference bugs, \Investigator had to survey the codebase for general vulnerability hypotheses on its own each iteration, increasing per-iteration cost by 2.1--19.9$\times$. In \sys, \Analyzer extracts seed-specific root causes once and passes them to \Investigator, which naturally spreads the search across different components and bug mechanisms. Without this, the agent fell back to broader searches that produced more overlapping scenarios, dropping the scenario coverage check pass rate by 17.9--21.3\%p (from 91.7--99.0\% down to 73.0--81.1\%).

\begin{table}[t]
    \centering
    \footnotesize
    \caption{Results without \Analyzer stage and reference bugs. \cc{\#} is the number of repeated iterations (without reference bugs) or seeds (\sys).}
    \label{tbl:reference-bugs}
    \input{tbl/reference-bugs}
\end{table}

\begin{table*}[t]
    \centering
    \footnotesize
    \caption{Bugs discovered by \sys in SpiderMonkey and JavaScriptCore. Ref. ID refers to the V8 seed that led to the discovery.}
    \label{tbl:cross-engine-bugs}
    \input{tbl/cross-engine-bugs}
\end{table*}

\subsection{RQ5. Cross-Engine Fuzzing}
\label{ss:eval-cross-engine}

\PP{Experiment setup}
To evaluate whether \sys can find bugs in other JavaScript engines using the seeds collected from V8, we ran \sys on SpiderMonkey and JavaScriptCore. We used the same four-stage pipeline with Claude Opus 4.6 and DPP-MAP scheduling. We ran each engine under a fixed API budget of \$1,000 and a fixed count of 100 seeds, following the same setup as in \autoref{ss:eval-pipeline}. However, due to the budget and time constraints, we ran all experiments only at the high effort level to maximize per-seed bug discovery as shown in \autoref{ss:eval-pipeline}.

\PP{Results}
\autoref{tbl:cross-engine-bugs} shows that reference bugs from one project can be used to find bugs in similar software. With 100 V8 seeds per engine, \sys found five bugs (including one duplicate) in SpiderMonkey and 14 in JavaScriptCore. \sys extracted root causes from V8 bugs, found relevant code in the other engines, and hypothesized how similar ideas could trigger bugs there. For example, seed 40057622 (CVE-2021-38001)~\cite{v8issue40057622} is a V8 bug where the property access cache mistakes which object it should read a property from, during \cc{super} property access. From this seed, \sys found JSC bug \#7 (Issue \jscbidVII), where JavaScriptCore's property access cache makes the same mistake in a different caching mechanism, reading a property value from a stale object instead of the intended one.

\subsection{Additional Experiments}
\label{ss:eval-additional}

\PP{Open Source Model}
We ran mini-experiments with two open-source models, Gemma 4~\cite{gemma4} and GLM-5.1~\cite{glm51}, to test whether they can be used in place of commercial models in \sys. As a result, we found that neither is yet competitive. Gemma 4 frequently produced malformed tool calls, causing the agent to fail before it could make meaningful progress; it found no bugs across 100 seeds. GLM-5.1 handled tool call formatting correctly but produced weaker hypotheses: it found only one bug after running 27 seeds, consuming approximately 700M tokens. These results suggest that open-source models are not yet suitable for agentic fuzzing at the level \sys requires, though we expect their performance to gradually improve in the future.

\PP{Generalization to Other Software}
Although our evaluation mainly targets JavaScript engines, the core techniques of \sys are not specific to them. To test this, we ran \sys on two additional targets outside our main evaluation. On ANGLE, Google Chrome's GPU graphics layer, \sys found one heap buffer overflow vulnerability~\cite{chromiumissue501476576}. We also found several vulnerabilities by adapting \sys to the Windows kernel, a closed-source target. Unfortunately, the Windows kernel lacks public bug details, so we had to manually craft seeds based on CVE entries. These results suggest that \sys can generalize beyond JavaScript engines and can even work on closed-source software when suitable seeds are provided. As future work, we plan to evaluate \sys on more targets and further explore its generalization capabilities.

%% file: tbl/pipeline.tex
\begin{tabular}{llrrrrrrr}
    \toprule
    & & & & & \multicolumn{2}{c}{\textbf{Cost (\$)}} & \multicolumn{2}{c}{\textbf{Time (min)}} \\
    \cmidrule(lr){6-7} \cmidrule(lr){8-9}
    \textbf{Effort} & \textbf{Pipeline} & \textbf{\#} & \textbf{TP} & \textbf{FP} & \textbf{Total} & \textbf{Avg.} & \textbf{Total} & \textbf{Avg.} \\
    \midrule
    \multicolumn{9}{l}{\textit{(a) Fixed budget (${\sim}$\$1{,}000)}} \\
    \multirow{3}{*}{High} & Four-stage & 39 & 3 & 1 & 986.3 & 25.3 & 5,081.3 & 130.3 \\
     & Three-stage & 62 & 3 & 2 & 994.0 & 16.0 & 5,121.9 & 82.6 \\
     & One-stage & 82 & \textbf{8} & 0 & 993.7 & 12.1 & 4,294.5 & 52.4 \\
    \addlinespace
    \multirow{3}{*}{Medium} & Four-stage & 53 & 2 & 2 & 995.5 & 18.8 & 4,269.1 & 80.5 \\
     & Three-stage & 92 & 4 & 0 & 993.6 & 10.8 & 4,122.8 & 44.8 \\
     & One-stage & 110 & \textbf{5} & 1 & 999.4 & 9.1 & 3,880.1 & 35.3 \\
    \addlinespace
    \multirow{3}{*}{Low} & Four-stage & 127 & \textbf{4} & 0 & 996.3 & 7.8 & 4,359.1 & 34.3 \\
     & Three-stage & 237 & 3 & 1 & 997.5 & 4.2 & 4,579.5 & 19.3 \\
     & One-stage & 300 & 2 & 2 & 998.5 & 3.3 & 3,654.9 & 12.2 \\
    \midrule
    \multicolumn{9}{l}{\textit{(b) Fixed seeds (100 each)}} \\
    \multirow{3}{*}{High} & Four-stage & 100 & \textbf{9} & 1 & 2,845.5 & 28.5 & 13,483.5 & 134.8 \\
     & Three-stage & 100 & 3 & 2 & 1,583.0 & 15.8 & 7,597.2 & 76.0 \\
     & One-stage & 100 & 8 & 1 & 1,188.6 & 11.9 & 5,070.7 & 50.7 \\
    \addlinespace
    \multirow{3}{*}{Medium} & Four-stage & 100 & 3 & 3 & 1,909.6 & 19.1 & 8,484.6 & 84.8 \\
     & Three-stage & 100 & \textbf{4} & 0 & 1,090.5 & 10.9 & 4,474.7 & 44.7 \\
     & One-stage & 100 & \textbf{4} & 1 & 917.9 & 9.2 & 3,507.3 & 35.1 \\
    \addlinespace
    \multirow{3}{*}{Low} & Four-stage & 100 & \textbf{4} & 0 & 766.2 & 7.7 & 3,305.0 & 33.0 \\
     & Three-stage & 100 & 3 & 0 & 409.0 & 4.1 & 1,826.0 & 18.3 \\
     & One-stage & 100 & 0 & 0 & 286.3 & 2.9 & 1,045.9 & 10.5 \\
    \bottomrule
\end{tabular}

%% file: tbl/reproducibility.tex
\begin{tabular}{l rr rr rr}
    \toprule
    & \multicolumn{2}{c}{\textbf{High}} & \multicolumn{2}{c}{\textbf{Medium}} & \multicolumn{2}{c}{\textbf{Low}} \\
    \cmidrule(lr){2-3} \cmidrule(lr){4-5} \cmidrule(lr){6-7}
    \textbf{Pipeline} & \textbf{pass@5} & \textbf{Total} & \textbf{pass@5} & \textbf{Total} & \textbf{pass@5} & \textbf{Total} \\
    \midrule
    Four-stage & \textbf{15}/21 & \textbf{32}/105 & 9/21 & \textbf{19}/105 & 4/21 & \textbf{12}/105 \\
    Three-stage & 8/21 & 15/105 & \textbf{11}/21 & \textbf{19}/105 & 5/21 & 8/105 \\
    One-stage & 13/21 & 25/105 & 9/21 & 17/105 & \textbf{6}/21 & 7/105 \\
    \bottomrule
\end{tabular}

%% file: tbl/seed-scheduling.tex
\begin{tabular}{lrrrrrrrr}
    \toprule
    & & & & & \multicolumn{2}{c}{\textbf{Cost (\$)}} & \multicolumn{2}{c}{\textbf{Time (min)}} \\
    \cmidrule(lr){6-7} \cmidrule(lr){8-9}
    \textbf{Strategy} & \textbf{\#} & \textbf{TP} & \textbf{FP} & \textbf{Cov.} & \textbf{Total} & \textbf{Avg.} & \textbf{Total} & \textbf{Avg.} \\
    \midrule
    \multicolumn{9}{l}{\textit{(a) Fixed budget (${\sim}$\$1{,}000)}} \\
    DPP-MAP & 53 & 2 & 2 & 99.0\% & 995.5 & 18.8 & 4,269.1 & 80.5 \\
    Newest & 44 & 1 & 1 & 83.2\% & 984.2 & 22.4 & 4,222.0 & 96.0 \\
    Random & 50 & 2 & 1 & 90.2\% & 973.8 & 19.5 & 4,456.4 & 89.1 \\
    \midrule
    \multicolumn{9}{l}{\textit{(b) Fixed seeds (100 each)}} \\
    DPP-MAP & 100 & 3 & 3 & 94.4\% & 1,909.6 & 19.1 & 8,484.6 & 84.8 \\
    Newest & 100 & 2 & 2 & 80.0\% & 2,260.7 & 22.6 & 10,275.7 & 102.8 \\
    Random & 100 & 6 & 2 & 90.7\% & 1,936.4 & 19.4 & 9,242.5 & 92.4 \\
    \bottomrule
\end{tabular}

%% file: tbl/reference-bugs.tex
\begin{tabular}{lrrrrrrrr}
    \toprule
    & & & & & \multicolumn{2}{c}{\textbf{Cost (\$)}} & \multicolumn{2}{c}{\textbf{Time (min)}} \\
    \cmidrule(lr){6-7} \cmidrule(lr){8-9}
    \textbf{Effort} & \textbf{\#} & \textbf{TP} & \textbf{FP} & \textbf{Cov.} & \textbf{Total} & \textbf{Avg.} & \textbf{Total} & \textbf{Avg.} \\
    \midrule
    \multicolumn{9}{l}{\emph{Without analyzer and reference bugs}} \\
    High$\ast$      & 2  & 1 & 2 & 76.4\% & 1,006.1 & 503.1 & 4,398.3 & 2,199.2 \\
    Medium$\ast$    & 20 & 1 & 2 & 81.1\% & 1,103.7 & 55.2  & 4,718.6 & 235.9   \\
    Low             & 61 & 0 & 0 & 73.0\% & 992.5   & 16.3  & 4,290.7 & 70.3    \\
    \midrule
    \multicolumn{9}{l}{\emph{With analyzer and reference bugs (\sys)}} \\
    High            & 39  & 3 & 1 & 97.7\% & 986.3   & 25.3  & 5,081.3 & 130.3   \\
    Medium          & 53  & 2 & 2 & 99.0\% & 995.5   & 18.8  & 4,269.1 & 80.5    \\
    Low             & 127 & 4 & 0 & 91.7\% & 996.3   & 7.8   & 4,359.1 & 34.3    \\
    \bottomrule
    \multicolumn{9}{l}{\footnotesize $\ast$ Includes one additional iteration that exceeded the \$1{,}000 budget.} \\
\end{tabular}

%% file: tbl/cross-engine-bugs.tex
\begin{tabular}{rllll}
    \toprule
    \textbf{\#} & \textbf{ID} & \textbf{Detail} & \textbf{Ref. ID} & \textbf{Status} \\
    \midrule
    \multicolumn{5}{l}{\textit{SpiderMonkey}} \\
    1 & \smbidI & Packed Field Narrowing/Widening Bypass in Wasm GC Struct Scalar Replacement & 40090201 & Duplicate \\
    2 & \smbidII & Missing canonicalizeValueZero in GenerateImportJitExit Externref Path & 40090201 & Reported \\
    3 & \smbidIII & \cc{JSOp::ArgumentsLength} returns wrong value for block-hoisted \cc{function arguments()\{\}} & 40089200 & Fixed \\
    4 & \smbidIV & \cc{for (using x of iterable)} in Generators --- Disposal Skipped on Generator Close & 40092882 & Reported \\
    5 & \smbidV & LOONG64 JIT [REDACTED] & 380604249 & Fixed \\
    \midrule
    \multicolumn{5}{l}{\textit{JavaScriptCore}} \\
    1 & \jscbidI & SIGSEGV in FTL [REDACTED] & 382547699 & Reported \\
    2 & \jscbidII & Hole-to-NaN Conversion in DFG Object Allocation Sinking Phase & 382547699 & Fixed \\
    3 & \jscbidIII & DFG [REDACTED] & 442086665 & Reported \\
    4 & \jscbidIV & DFG \cc{operationStringProtoFuncReplaceAllGeneric} Skips Global Flag Check for RegExp & 442086665 & Reported \\
    5 & \jscbidV & Array ToPrimitive Fast Path Ignores Object.prototype.valueOf Override & 406871259 & Reported \\
    6 & \jscbidVI & Incorrect Scope Resolution for Sloppy Function Hoisting from eval() in Default Parameter Expressions & 40089200 & Reported \\
    7 & \jscbidVII & Incorrect Property Read in Megamorphic Cache due to ownProperty Flag Confusion & 40057622 & Fixed \\
    8 & \jscbidVIII & LLInt/BBQ JIT Discrepancy in WebAssembly div/rem Constant Folding & 40055671 & Fixed \\
    9 & \jscbidIX & Unchecked [REDACTED] & 40064614 & Reported \\
    10 & \jscbidX & Set Spread in DFG/FTL Missing Per-Instance Prototype Check & 385386138 & Reported \\
    11 & \jscbidXI & Incorrect Math.round Result via floor(x+0.5) JIT Fast Path & 416128193 & Reported \\
    12 & \jscbidXII & YARR Omits Named Group from \cc{indices.groups} via Tracking-Slot Reset on Backtrack & 40093076 & Reported \\
    13 & \jscbidXIII & BBQ JIT Constant Folding Signed Zero Mishandling in \cc{computeFloatingPointMinOrMax} & 40070548 & Fixed \\
    14 & \jscbidXIV & Input Position Corruption in Yarr Backreference Backward Matching via Surrogate Pair Rewind & 40053946 & Reported \\
    \bottomrule
\end{tabular}

%% file: problem.tex
\section{Open Problems}
\label{s:problem}

Despite its effectiveness, agentic fuzzing is still in its early stages. There are many open problems that need to be addressed.

\PP{Cost effectiveness}
Cost effectiveness remains a major challenge for agentic fuzzing. \sys is still expensive to run. As a result, we could evaluate only about 23.8\% of the collected seeds from the Chromium issue tracker~\cite{chromiumissuetracker} (see \autoref{ss:eval-newbugs}). This limitation also prevents us from evaluating \sys on more projects. While we attempted to mitigate this limitation by introducing techniques like scenario coverage (see \autoref{sss:scenario-coverage}) and seed scheduling (see \autoref{sss:seed-scheduling}), we believe that more research is needed to further improve the cost efficiency of agentic fuzzing while maintaining its effectiveness.

\PP{No reference bug}
\sys assumes that reference bugs are available; however, this is not always the case. Reference bugs guide the agent toward specific components and root causes, as shown in \autoref{ss:eval-reference-bugs}: without them, the agent falls back to broad, undirected searches that cost more and find fewer bugs. For well-studied open-source targets like V8, years of public bug reports with detailed root causes and PoCs provide a rich seed corpus.

However, not all projects provide such references. Closed-source software like Windows only publishes CVE entries with short descriptions (e.g., ``Windows kernel buffer overflow''), which lack the detail \sys can refer to. Newer or less-documented open-source projects may simply have too few prior reports. In these settings, \sys cannot be directly applied. One possible direction is to bootstrap seeds from related projects, as our cross-engine experiments (\autoref{ss:eval-cross-engine}) suggest that root causes can transfer across implementations that share architectural similarities. Another direction is to let the agent build its own seed corpus from documentation, changelogs, or code review history, though this remains unexplored.

\PP{Design uncertainty}
Design choices for agentic fuzzing are hard to justify definitively. Our pipeline comparison (\autoref{ss:eval-pipeline}) demonstrates this: the four-stage pipeline achieved the highest per-seed bug discovery (9 bugs from 100 seeds at high effort), but the one-stage pipeline is competitive (8 bugs with the same setup) because the agent self-organizes using subagents and task management on its own. This aligns with recent findings that single-agent systems can outperform multi-agent designs under equal token budgets~\cite{tran2026single}. We cannot confidently say which design is better---it depends on budget, model, and target. More broadly, \sys's design is based on empirical observations about current LLMs, not on well-understood principles. This makes the design inherently heuristic and fragile across model generations. For example, few-shot prompting~\cite{brown2020fewshot} improved performance in earlier non-reasoning models, but the effect is reduced in reasoning models~\cite{guo2025deepseek}. A design choice that works today may not survive the next model update.

%% file: discussion.tex
\section{Discussion}
\label{s:discussion}

\PP{Bug Explainability}
Compared to traditional fuzzing or static analysis tools, \sys provides an advantage in explaining bugs. Traditional tools typically report a crash or a violation of an assertion, which may not provide much insight into the root cause of the bug. In contrast, \sys generates detailed reports that include the steps taken to discover the bug, the reasoning behind each step, and the potential impact of the bug. This level of detail can help developers understand the underlying issue more quickly and accurately, leading to faster and more effective fixes. For example, when we reported a duplicate bug (Issue~\bidXI), which was found by the fuzzers of the V8 team a day before our report, a V8 developer commented ``Our fuzzers didn't provide an explanation of what's happening, so thanks for that!''. This feedback suggests that \sys's explanations can be valuable even for bugs that are already known, as they can help developers save time and effort in diagnosing and fixing the issue.

\PP{Implications}
Our results imply that current commercial LLMs, with proper harnessing, can discover vulnerabilities that skilled human auditors can find. This raises a concern: threat actors can use the same techniques to find vulnerabilities that were previously out of reach and exploit them for real attacks. At the same time, the defenders' side is already under pressure. Bug report volume is growing, and AI-generated slop reports (e.g., \cite{v8issue491749534}) add noise that developers must still triage. Developers increasingly turn to AI-based techniques such as Automatic Program Repair (APR) to keep up, but APR still requires human review to ensure correctness. The result is an asymmetry: attackers can scale discovery through automation, while defenders remain bottlenecked by human verification. This imbalance disproportionately affects low-resource entities that lack the budget to absorb the cost of either AI-assisted defense or the increased volume of vulnerabilities.

%% file: relwk.tex
\section{Related Work}
\label{s:relwk}

\PP{LLM-based vulnerability discovery}
Researchers have recently been using LLM agents to discover vulnerabilities. Big Sleep~\cite{bigsleep2024} discovered several impactful vulnerabilities in multiple open-source projects, including SQLite~\cite{sqlite} and Chromium~\cite{chromium}, using LLM agents in variant analysis. {\AE}SIR~\cite{aesir2026} also found 21 CVEs in AI infrastructure by performing variant analysis of new CVEs. XBOW~\cite{demoor2024xbow}, PentestGPT~\cite{deng2024pentestgpt}, and other works~\cite{happe2023pwn,shen2025pentestagent,david2025multi} automate penetration testing with LLM agents by giving them tools to dynamically test the target system, while IRIS~\cite{li2025iris} and LLMxCPG~\cite{lekssays2025llmxcpg} combine LLMs with static analysis for vulnerability detection.
Similar to Big Sleep, \sys discovers new bugs from reference bugs using LLM agents. However, we disclose \sys's architecture and details while Big Sleep does not, and we also introduce several optimizations that can make \sys more cost-effective for general users. Moreover, \sys found new bugs in the same V8 environment where Big Sleep had already found vulnerabilities.

\PP{LLM-assisted fuzzing}
A line of work has focused on using LLMs to assist fuzzing. Several works~\cite{deng2023titanfuzz,deng2024fuzzgpt,xia2024fuzz4all,yang2024whitefox,yang2025hybrid} use LLMs to generate or mutate test inputs for language processors or deep learning libraries. ELFuzz~\cite{chen2025elfuzz} and G2Fuzz~\cite{zhang2025g2fuzz} extend this by generating and mutating input generators rather than test inputs. ChatAFL~\cite{meng2024chatafl} leverages LLMs' knowledge of protocol specifications to guide protocol fuzzing, while KernelGPT~\cite{yang2025kernelgpt} uses LLMs to generate syzkaller specifications for kernel fuzzing. PromptFuzz~\cite{lyu2024promptfuzz} uses LLMs to iteratively fuzz LLM prompts to generate fuzzing harnesses. CovRL~\cite{eom2024covrl} combines LLM-based mutation with fine-tuning LLM from coverage feedback in JavaScript engine fuzzing. While these works focus on augmenting fuzzing, \sys uses deep agents to directly reason about code semantics to find bugs.

\PP{Variant analysis}
Traditionally, variant analysis found bugs by matching code patterns or signatures. CodeQL~\cite{codeql} and Joern~\cite{yamaguchi2014modeling} provide a query language to define buggy code patterns and allow these patterns to be matched against codebases. VUDDY~\cite{kim2017vuddy}, TRACER~\cite{kang2022tracer}, and others~\cite{jang2012redebug,woo2022movery,xiao2020mvp,woo2023v1scan,feng2024fire,huang2024vmud,xu2025enhancing} detect variants by matching abstracted code signatures, code hashes, program slices, or traces. These approaches rely on syntactic or structural similarity, which limits their ability to find logic bug variants that share semantic patterns but differ in code structure.

\PP{JavaScript engine fuzzing}
Fuzzing has been the predominant technique used to find bugs in JavaScript engines. JavaScript engine fuzzing spans \WC{1} grammar-based fuzzing~\cite{holler2012langfuzz,wang2017skyfire,aschermann2019nautilus,wang2019superion}, \WC{2} semantics-aware generation~\cite{han2019codealchemist,wong2025temujs}, \WC{3} language model-based generation~\cite{lee2020montage, eom2024covrl}, \WC{4} mutation-based fuzzing~\cite{he2021sofi,gross2023fuzzilli,xu2024fuzzflow,wang2024optfuzz,wang2025bcfuzz}, \WC{5} differential and oracle-based testing~\cite{bernhard2022jitpicking,wang2023fuzzjit,wachter2025dumpling}, \WC{6} binding-layer and cross-language fuzzing~\cite{dinh2021favocado,lin2026madeye}, and \WC{7} conformance testing~\cite{park2021jest,ye2021comfort}. While these fuzzers have shown their effectiveness in finding bugs, they still struggle to find logic bugs that require semantic understanding.

%% file: conclusion.tex
\section{Conclusion}
\label{s:conclusion}

We propose \emph{agentic fuzzing}, which takes reference bugs as seeds and uses deep agents to find similar bugs through code reasoning. We designed an agentic fuzzer \sys with a four-stage agent pipeline, scenario coverage to avoid redundant investigations across seeds, and DPP-MAP seed scheduling to cover diverse root causes early. We evaluated \sys primarily on the V8 JavaScript engine, where it found \bugtotal bugs (including \bugduplicate duplicates) with a total bounty of \$35{,}000 and two CVEs (\cveII and \cveVIII). Several of these are complex logic bugs that conventional fuzzing and static analysis struggle to find. We further showed that V8 seeds can be used in other engines, finding five bugs in SpiderMonkey (including one duplicate) and 14 in JavaScriptCore. However, agentic fuzzing is still in its early stages, with open problems including cost effectiveness, the need for reference bugs, and design uncertainty.

%% file: appendix.tex
\appendix

\label{s:appendix}
\label{s:appendix:asdf}

\section{Ethical Considerations}
\label{s:ethics}

As shown in \autoref{tbl:bugs} and \autoref{tbl:cross-engine-bugs}, all bugs found during our research, including those found in \autoref{ss:eval-additional}, were reported to the respective vendors (e.g., Google). We redacted some bug details to avoid public disclosure before each vendor's disclosure deadline --- 90 days after a fix is released.

\section{Generative AI Usage}
\label{s:aiusage}

We used generative AI in two ways: as the core component of \sys and as a development aid. First, \sys runs on Claude models, primarily Claude Opus 4.6~\cite{opus46}. As described in \autoref{s:impl}, the agents use Claude Opus 4.6 through the Claude Agent SDK~\cite{claudeagentsdk}, and the scenario coverage check (\autoref{ss:design-exploration}) uses a lighter Claude model (i.e., Haiku) for semantic comparison. We also used open-source models gemma4~\cite{gemma4} and GLM 5.1~\cite{glm51} for evaluation (\autoref{ss:eval-additional}).
Therefore, all bug reports of \sys are produced by generative AI models. We independently checked every bug report by reviewing it, reproducing the PoC, and confirming the claimed behaviors before submitting the report to the respective vendor.

Second, we used Claude Code~\cite{claudecode} with the Opus 4.6 model~\cite{opus46} to quickly prototype and develop \sys, including the evaluation variants (e.g., the different pipeline designs). We thoroughly supervised the entire development process, and we conducted all experiments and checked all their results manually.

\section{Example of \sys Finding a Bug}
\label{s:appendix:design_example}

Using the motivating example (\autoref{ss:motiv-variant}), we walk through how the pipeline discovers the Turboshaft variant from the Maglev reference bug.

Given the reference bug CVE-2025-10892~\cite{referencebug} as a URL, \Analyzer starts by fetching the issue contents, including the discussion, patches, and PoC code. \Analyzer then dispatches \cc{Explore} subagents to read relevant source files such as \cc{maglev-ir.h}, \cc{maglev-graph-builder.cc}, and \cc{base/bit-field.h}. After reading these files, \Analyzer determines that the root cause of the reference bug is an integer overflow in a compiler IR metadata field (i.e., \cc{size\_t input\_count} in \cc{Operation}) that is only guarded by a debug assertion. \Analyzer captures this insight as the \emph{bug mechanism} of the reference bug, which is a concise description of the root cause that later stages can use to search for variants.

\Investigator takes the report from \Analyzer and scans the V8 codebase for similar root causes. It starts with Maglev, where the reference bug was found, and searches for other integer fields in Maglev's IR classes that are guarded only by debug assertions. After thoroughly searching Maglev code, \Investigator widens the search to other backends with grep queries such as \cc{BitField64.*Count} and \cc{uint16\_t.*input\_count}, and lands on Turboshaft, which shares a similar mechanism with Maglev. In Turboshaft, the \cc{Operation} constructor (\cc{operations.h}) narrows an incoming \cc{size\_t} input count to a \cc{uint16\_t} and guards it with a \cc{DCHECK}. \Investigator traces the construction path from the start of graph building to the \cc{PhiOp} construction, and notices that \cc{PhiOp} does not have an explicit pre-construction check. \Investigator hypothesizes that a \cc{PhiOp} with more than 65,535 predecessors would cause the count to be silently truncated, leading to memory corruption, and submits a scenario describing the trigger conditions, affected source locations, and an advisory note pointing to \cc{br\_table} as a way to inflate the predecessor count.

\ScenarioAnalyzer takes the scenario from \Investigator and writes a PoC to trigger the vulnerability. It reads the relevant source files such as \cc{turboshaft-graph-interface.cc} and \cc{ReduceSwitch}, and learns that each \cc{br\_if} or \cc{br\_table} target adds one predecessor at the destination block. It also finds the per-table cap \cc{kV8MaxWasmFunctionBrTableSize\,=\,65{,}520} by reading \cc{wasm-limits.h}. Thinking that a single \cc{br\_table} cannot exceed the 16-bit limit, it decides to use two \cc{br\_table} instructions in opposite arms of an if-else, each with 33,000 entries targeting the same outer block, to trigger the overflow (shown in \autoref{fig:motiv-trigger}). \ScenarioAnalyzer then initially runs the PoC on a release build, which completes without any observable evidence because the integer truncation silently occurs. \ScenarioAnalyzer then re-runs the same PoC on a debug build, where the \cc{DCHECK} on \cc{operations.h} fires, confirming that the truncation occurs and violates the invariant defined by the \cc{DCHECK}.

\Validator independently verifies the bug by re-executing the PoC on a clean V8 build and observing the same release and debug results. It confirms that the debug assertion failure matches the hypothesized integer truncation rather than an unrelated one, and that no security-disabling flags or debug-only natives are used in reproducing the bug. After verification, \Validator accepts the scenario and writes a vulnerability report covering summary, trigger conditions, reproduction steps with both release and debug outputs, and a suggested patch with an input-count check before \cc{PhiOp} construction.

\section{Case Studies}
\label{s:appendix:case-studies}

\PP{Bug \#9: parseInt exponent overflow}
V8 has a bug in the JavaScript \cc{parseInt} function (converts strings to integers) that makes the function return \cc{0} instead of \cc{Infinity} when parsing long strings with a power-of-two radix (e.g., 32). When V8 parses a long string whose numeric value exceeds 53-bit precision, it truncates the low bits of the integer and tracks the excess as an integer exponent. The problem is that for radix 32, the exponent can grow large enough to overflow a signed integer when a string of more than $429{,}496{,}740$ characters is parsed (e.g., \cc{parseInt("1".repeat(429496741), 32)}). The overflow makes the exponent negative, so instead of returning \cc{Infinity} as it should, \cc{parseInt} incorrectly returns \cc{0}. This bug has existed since March 2010~\cite{v8issue1374005}, for over 16 years.

\sys could recognize the tricky trigger conditions of this bug by analyzing the code and reasoning about how the exponent grows with string length and radix. Because the bug requires a string of over 429 million characters, it needs 64-bit builds to trigger, since V8 limits string length to about 512MB (i.e., $(1 << 29) - 24$) on 64-bit builds and about 256MB (i.e., $(1 << 28) - 24$) on 32-bit builds. Among power-of-2 radixes, only the maximum radix 32 has an overflow threshold that fits under the string length limit, so the bug is specific to radix 32. Because of these specific and non-obvious trigger conditions, this bug has survived for over 16 years without being discovered by fuzzing or manual auditing, demonstrating the effectiveness of \sys in finding hard-to-find bugs that require deep understanding and complex reasoning.

\begin{figure}[t]
    \resizebox{\columnwidth}{!}{
        \begin{minipage}{\columnwidth}
            \input{code/eval-bug12-poc.js}
        \end{minipage}
    }
    \caption{
        Minified PoC for bug \#12. The loop update \cc{i = (3 - i) | 0} oscillates \cc{i} between 2 and 1, so \cc{count} reaches 5. After loop peeling, which extracts the first iteration and makes $i=3-2=1$, Turboshaft simulates the update as \cc{i - 3} instead of \cc{3 - i}, predicts the loop exits after 1 iteration, and unrolls it to return 2.
    }
    \label{fig:eval-bug12-poc}
\end{figure}

\PP{Bug \#12: Turboshaft loop unrolling}
Turboshaft had a bug in its loop unrolling optimization that caused incorrect calculations of iteration counts for loops that contain subtractions. Before unrolling a loop, Turboshaft simulates the loop execution to compute how many iterations it will run. However, its simulation logic wrongly assumes that the subtraction operator is commutative, making it wrongly compute loop variable updates of \cc{i = c - i} as if they were \cc{i = i - c}. As a result, Turboshaft miscomputes the loop variable's value and predicts the wrong iteration count, which can lead to under-unrolling and miscompilation.

\sys managed to construct a non-trivial oscillating loop to trigger this bug, shown in \autoref{fig:eval-bug12-poc}. The loop variable \cc{i} is updated as \cc{i = (3 - i) | 0}, which oscillates \cc{i} between 2 and 1. The loop body increments \cc{count} and breaks when \cc{count >= 5}, so the loop runs for 5 iterations before breaking. However, in the case of Turboshaft, after the loop peeling phase extracts the first iteration with \cc{i=2}, the remaining loop has the update \cc{i = 3 - i} with an initial value \cc{i=1}. Turboshaft then simulates this update as if it were \cc{i = i - 3}, which produces the sequence $1 \to 1-3 = -2$, and predicts the loop exits after just 1 iteration. Therefore, it unrolls the loop into only two iterations, making the optimized code return 2 instead of 5.

\begin{table*}[t]
    \centering
    \footnotesize
    \caption{Detailed bug reproducibility across pipeline design choices. Each cell shows the number of times (out of 5 runs) a bug was discovered by the given design. Bugs that correspond to entries in \autoref{tbl:bugs} are annotated with their bug number and ID. Three bugs (marked --) were independently fixed before we reported them and are therefore not listed in \autoref{tbl:bugs}.}
    \label{tbl:reproducibility-full}
    \resizebox{\textwidth}{!}{
        \input{tbl/reproducibility-full}
    }
\end{table*}

%% file: code/eval-bug12-poc.js.tex
\begin{Verbatim}[commandchars=\\\{\},codes={\catcode`\$=3\catcode`\^=7\catcode`\_=8\relax}]
\PY{k+kd}{function}\PY{+w}{ }\PY{n+nx}{f}\PY{p}{(}\PY{p}{)}\PY{+w}{ }\PY{p}{\PYZob{}}
\PY{+w}{  }\PY{k+kd}{let}\PY{+w}{ }\PY{n+nx}{count}\PY{+w}{ }\PY{o}{=}\PY{+w}{ }\PY{l+m+mf}{0}\PY{p}{;}
\PY{+w}{  }\PY{k}{for}\PY{+w}{ }\PY{p}{(}\PY{k+kd}{let}\PY{+w}{ }\PY{n+nx}{i}\PY{+w}{ }\PY{o}{=}\PY{+w}{ }\PY{l+m+mf}{2}\PY{p}{;}\PY{+w}{ }\PY{n+nx}{i}\PY{+w}{ }\PY{o}{\PYZgt{}=}\PY{+w}{ }\PY{l+m+mf}{0}\PY{p}{;}\PY{+w}{ }\PY{n+nx}{i}\PY{+w}{ }\PY{o}{=}\PY{+w}{ }\PY{p}{(}\PY{l+m+mf}{3}\PY{+w}{ }\PY{o}{\PYZhy{}}\PY{+w}{ }\PY{n+nx}{i}\PY{p}{)}\PY{+w}{ }\PY{o}{|}\PY{+w}{ }\PY{l+m+mf}{0}\PY{p}{)}\PY{+w}{ }\PY{p}{\PYZob{}}
\PY{+w}{    }\PY{c+c1}{// i oscillates: 2 \PYZhy{}\PYZgt{} 1 \PYZhy{}\PYZgt{} 2 \PYZhy{}\PYZgt{} 1 \PYZhy{}\PYZgt{} ...}
\PY{+w}{    }\PY{n+nx}{count}\PY{+w}{ }\PY{o}{=}\PY{+w}{ }\PY{p}{(}\PY{n+nx}{count}\PY{+w}{ }\PY{o}{+}\PY{+w}{ }\PY{l+m+mf}{1}\PY{p}{)}\PY{+w}{ }\PY{o}{|}\PY{+w}{ }\PY{l+m+mf}{0}\PY{p}{;}
\PY{+w}{    }\PY{k}{if}\PY{+w}{ }\PY{p}{(}\PY{n+nx}{count}\PY{+w}{ }\PY{o}{\PYZgt{}=}\PY{+w}{ }\PY{l+m+mf}{5}\PY{p}{)}\PY{+w}{ }\PY{k}{break}\PY{p}{;}
\PY{+w}{  }\PY{p}{\PYZcb{}}
\PY{+w}{  }\PY{k}{return}\PY{+w}{ }\PY{n+nx}{count}\PY{p}{;}\PY{+w}{ }\PY{c+c1}{// should be 5}
\PY{p}{\PYZcb{}}
\PY{o}{\PYZpc{}}\PY{n+nx}{PrepareFunctionForOptimization}\PY{p}{(}\PY{n+nx}{f}\PY{p}{)}\PY{p}{;}
\PY{n+nx}{f}\PY{p}{(}\PY{p}{)}\PY{p}{;}
\PY{o}{\PYZpc{}}\PY{n+nx}{OptimizeFunctionOnNextCall}\PY{p}{(}\PY{n+nx}{f}\PY{p}{)}\PY{p}{;}
\PY{n+nx}{f}\PY{p}{(}\PY{p}{)}\PY{p}{;}\PY{+w}{ }\PY{c+c1}{// interpreter: 5 (correct), Turboshaft: 2 (wrong)}
\end{Verbatim}

%% file: tbl/reproducibility-full.tex
\begin{tabular}{r r l rrr rrr rrr}
    \toprule
    & & & \multicolumn{3}{c}{\textbf{High}} & \multicolumn{3}{c}{\textbf{Medium}} & \multicolumn{3}{c}{\textbf{Low}} \\
    \cmidrule(lr){4-6} \cmidrule(lr){7-9} \cmidrule(lr){10-12}
    \textbf{\#} & \textbf{ID} & \textbf{Detail} & \textbf{Four} & \textbf{Three} & \textbf{One} & \textbf{Four} & \textbf{Three} & \textbf{One} & \textbf{Four} & \textbf{Three} & \textbf{One} \\
    \midrule
    -- & -- & Maglev Phi Representation Selector Type Confusion & \cellcolor{green!20}1/5 & 0/5 & 0/5 & 0/5 & 0/5 & 0/5 & 0/5 & 0/5 & 0/5 \\
    13 & \bidXIII & ARM32 Assembler ClearInternalState Invariant & \cellcolor{green!100}5/5 & \cellcolor{green!60}3/5 & \cellcolor{green!80}4/5 & \cellcolor{green!100}5/5 & \cellcolor{green!60}3/5 & \cellcolor{green!80}4/5 & \cellcolor{green!80}4/5 & \cellcolor{green!40}2/5 & 0/5 \\
    23 & \bidXXIII & Missing Immutability Check for \cc{using}/\cc{await using} Exports & \cellcolor{green!60}3/5 & \cellcolor{green!20}1/5 & \cellcolor{green!20}1/5 & \cellcolor{green!20}1/5 & 0/5 & 0/5 & 0/5 & 0/5 & 0/5 \\
    -- & -- & Stack Overflow in Generate\_ResumeGeneratorTrampoline & \cellcolor{green!100}5/5 & \cellcolor{green!100}5/5 & \cellcolor{green!60}3/5 & \cellcolor{green!40}2/5 & \cellcolor{green!60}3/5 & \cellcolor{green!40}2/5 & \cellcolor{green!100}5/5 & \cellcolor{green!20}1/5 & \cellcolor{green!20}1/5 \\
    26 & \bidXXVI & Fast API kUint64 Signed/Unsigned Confusion & \cellcolor{green!20}1/5 & 0/5 & 0/5 & 0/5 & 0/5 & 0/5 & 0/5 & 0/5 & 0/5 \\
    24 & \bidXXIV & Missing PrecededByMember Check in ParseReturnStatement & \cellcolor{green!80}4/5 & \cellcolor{green!20}1/5 & \cellcolor{green!20}1/5 & \cellcolor{green!60}3/5 & \cellcolor{green!60}3/5 & \cellcolor{green!20}1/5 & \cellcolor{green!20}1/5 & 0/5 & 0/5 \\
    27 & \bidXXVII & Private Name Resolution via FinalizeBlockScope Skip Bit Loss & \cellcolor{green!20}1/5 & 0/5 & 0/5 & 0/5 & 0/5 & 0/5 & 0/5 & 0/5 & 0/5 \\
    7 & \bidVII & TurboFan SpeculativeAdditiveSafeIntegerAdd Type Narrowing & \cellcolor{green!20}1/5 & 0/5 & 0/5 & 0/5 & 0/5 & 0/5 & 0/5 & 0/5 & 0/5 \\
    28 & \bidXXVIII & Deoptimizer Undefined NaN Constant Mishandling & \cellcolor{green!20}1/5 & 0/5 & 0/5 & 0/5 & 0/5 & 0/5 & 0/5 & 0/5 & 0/5 \\
    38 & \bidXXXVIII & Immutability Bypass in JIT DataView Setters & \cellcolor{green!60}3/5 & 0/5 & \cellcolor{green!20}1/5 & \cellcolor{green!20}1/5 & \cellcolor{green!20}1/5 & 0/5 & 0/5 & 0/5 & 0/5 \\
    21 & \bidXXI & Missing Escape Sequence Check for \cc{using} Keyword & \cellcolor{green!20}1/5 & \cellcolor{green!40}2/5 & \cellcolor{green!60}3/5 & \cellcolor{green!40}2/5 & \cellcolor{green!60}3/5 & \cellcolor{green!20}1/5 & \cellcolor{green!40}2/5 & 0/5 & \cellcolor{green!20}1/5 \\
    22 & \bidXXII & Double-Dispose of \cc{using} in C-style For-Loop & \cellcolor{green!20}1/5 & \cellcolor{green!20}1/5 & \cellcolor{green!60}3/5 & 0/5 & \cellcolor{green!20}1/5 & \cellcolor{green!80}4/5 & 0/5 & \cellcolor{green!40}2/5 & \cellcolor{green!20}1/5 \\
    14 & \bidXIV & Wrong DCHECK in DependOnContextCell & 0/5 & \cellcolor{green!20}1/5 & \cellcolor{green!20}1/5 & 0/5 & \cellcolor{green!20}1/5 & 0/5 & 0/5 & 0/5 & \cellcolor{green!20}1/5 \\
    20 & \bidXX & Missing Prototype Chain Dependency for ModuleExport & \cellcolor{green!40}2/5 & 0/5 & 0/5 & \cellcolor{green!20}1/5 & \cellcolor{green!20}1/5 & \cellcolor{green!20}1/5 & 0/5 & 0/5 & 0/5 \\
    12 & \bidXII & Non-Commutative Subtraction in Turboshaft Loop Unrolling & \cellcolor{green!40}2/5 & \cellcolor{green!20}1/5 & \cellcolor{green!80}4/5 & 0/5 & 0/5 & \cellcolor{green!40}2/5 & 0/5 & 0/5 & 0/5 \\
    32 & \bidXXXII & IterableForEach Missing Array Prototype Check & 0/5 & 0/5 & \cellcolor{green!20}1/5 & \cellcolor{green!40}2/5 & \cellcolor{green!20}1/5 & 0/5 & 0/5 & \cellcolor{green!20}1/5 & 0/5 \\
    31 & \bidXXXI & RISCV64 Maglev CompareIntPtrAndBranch Codegen & 0/5 & 0/5 & \cellcolor{green!20}1/5 & 0/5 & 0/5 & 0/5 & 0/5 & 0/5 & 0/5 \\
    30 & \bidXXX & RISCV64 Maglev ToUint8Clamped Codegen & 0/5 & 0/5 & \cellcolor{green!20}1/5 & 0/5 & 0/5 & 0/5 & 0/5 & 0/5 & 0/5 \\
    25 & \bidXXV & JSBuiltinsConstructStub Stack Overflow on RISCV/LOONG64/MIPS64 & \cellcolor{green!20}1/5 & 0/5 & 0/5 & \cellcolor{green!40}2/5 & \cellcolor{green!20}1/5 & 0/5 & 0/5 & \cellcolor{green!40}2/5 & \cellcolor{green!40}2/5 \\
    -- & -- & LOONG64 Maglev Branch Offset Overflow & 0/5 & 0/5 & 0/5 & 0/5 & \cellcolor{green!20}1/5 & \cellcolor{green!20}1/5 & 0/5 & 0/5 & 0/5 \\
    17 & \bidXVII & LOONG64 PatchToJump SIGSEGV via Missing RwxMemoryWriteScope & 0/5 & 0/5 & \cellcolor{green!20}1/5 & 0/5 & 0/5 & \cellcolor{green!20}1/5 & 0/5 & 0/5 & \cellcolor{green!20}1/5 \\
    \midrule
    & & pass@5 & \cellcolor{green!71}15/21 & \cellcolor{green!38}8/21 & \cellcolor{green!62}13/21 & \cellcolor{green!43}9/21 & \cellcolor{green!52}11/21 & \cellcolor{green!43}9/21 & \cellcolor{green!19}4/21 & \cellcolor{green!24}5/21 & \cellcolor{green!29}6/21 \\
    & & Total & \cellcolor{green!30}32/105 & \cellcolor{green!14}15/105 & \cellcolor{green!24}25/105 & \cellcolor{green!18}19/105 & \cellcolor{green!18}19/105 & \cellcolor{green!16}17/105 & \cellcolor{green!11}12/105 & \cellcolor{green!8}8/105 & \cellcolor{green!7}7/105 \\
    \bottomrule
\end{tabular}